\documentclass[12pt]{article}
\usepackage{soul}
\usepackage{hyperref}
\usepackage{cite}
\usepackage{color}
\usepackage{graphicx}
\usepackage{amsmath}
\usepackage{amssymb}
\usepackage{xspace}
\usepackage{verbatim}
\usepackage{mathtools}

\DeclarePairedDelimiter\floor{\lfloor}{\rfloor}

\makeatletter
\@addtoreset{equation}{section}

\makeatletter
\renewcommand\section{\@startsection {section}{1}{\z@}%
                                   {-3.5ex \@plus -1ex \@minus -.2ex}
                                   {2.3ex \@plus.2ex}%
                                   {\normalfont\large\bfseries}}
\renewcommand\subsection{\@startsection{subsection}{2}{\z@}%
                                     {-3.25ex\@plus -1ex \@minus -.2ex}%
                                     {1.5ex \@plus .2ex}%
                                     {\normalfont\bfseries}}

\def\baselinestretch{1.2}
\newcommand{\be}{\begin{equation}}
\newcommand{\ee}{\end{equation}}
\newcommand{\beq}{\begin{eqnarray}}
\newcommand{\eeq}{\end{eqnarray}}

\newcommand{\gone}[1]{{}}

\begin{document}
\begin{titlepage}
\begin{flushright}
MAD-TH-16-02
\end{flushright}

\vfil

\begin{center}

{\bf \Large
Resolved gravity duals of ${\cal N}=4$ quiver field theories\\ in 2+1
dimensions
}

\vfil

William Cottrell and Akikazu Hashimoto

\vfil

Department of Physics, University of Wisconsin, Madison, WI 53706, USA

\vfil

\end{center}

\begin{abstract}
\noindent We generalize the construction by Aharony, Hashimoto,
Hirano, and Ouyang of ${\cal N}=4$ quiver gauge theory with gauge
group $U(N+M) \times U(N)$, $k$ fundamentals charged under $U(N)$ and
bi-fundamentals, to the case with gauge group $\prod_{i=1}^{\hat k}
U(N_i)$ with $k_i$ fundamentals charged under $U(N_i)$. This
construction is facilitated by considering the resolved $ALE_{\hat k}
\times TN_{k}$ background in M-theory including non-trivial fluxes
through the resolved 4-cycles in the geometry. We also describe the
M-theory lift of the IIA Page charge quantization condition. Finally, we
clarify the role of string corrections in various regimes of parameter
space.

\end{abstract}
\vspace{0.5in}

\end{titlepage}
\renewcommand{\baselinestretch}{1.05}  

\section{Introduction}

${\cal N}=4$ gauge field theories in $2+1$ dimensions is a rich
dynamical system, exhibiting features such as quantum corrected moduli
spaces and enhancons in the holographic dual. The number of
supersymmetries is the same as theories in 3+1 dimensions with ${\cal
  N}=2$ supersymmetry, whose vacuum structure can be analyzed exactly
using the techniques of Seiberg and Witten
\cite{Seiberg:1994rs,Seiberg:1994aj}.  The relation between ${\cal
  N}=2$ theories in $3+1$ dimensions and ${\cal N}=4$ theories in
$2+1$ dimensions is a rich subject on its own
\cite{Seiberg:1996nz}. One complication stems from the fact that
${\cal N}=4$ vector multiplets in $2+1$ dimensions have twice as many
scalar components compared to ${\cal N}=2$ vector multipletes in $3+1$
dimensions. There are, however, other powerful tools at our disposal
to explore the quantum dynamics of ${\cal N}=4$ theories in 2+1
dimensions, such as mirror symmetry \cite{Intriligator:1996ex},
localization, and holography. Recently, the structure of Coulomb
branch of ${\cal N}=4$ theories in 2+1 dimensions was mapped out
formally in \cite{Bullimore:2015lsa}.

The fact that full quantum dynamics is accessible on the field theory
side provides an opportunity to explore subtle issues in gauge/gravity
duality where string and quantum corrections are expected on the
gravity side of the correspondence. Knowing the existence, and in some
instance, the analytic form, of certain physical quantities on the
field theory side provides a concrete target that one can aim to
reproduce on gravity side.

\begin{figure}
\centerline{\begin{tabular}{cc}
\includegraphics{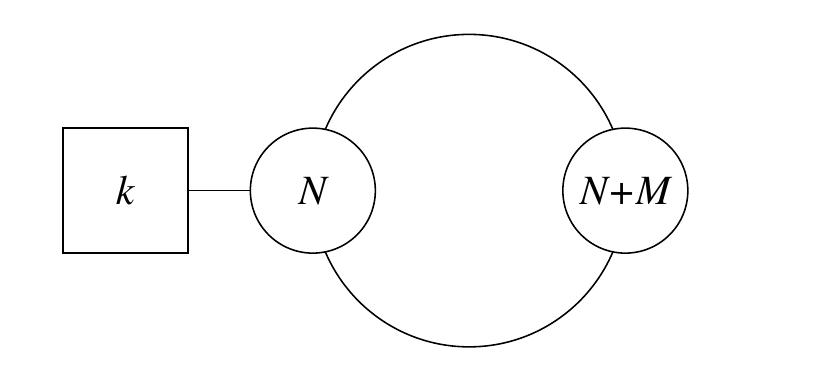} & \includegraphics{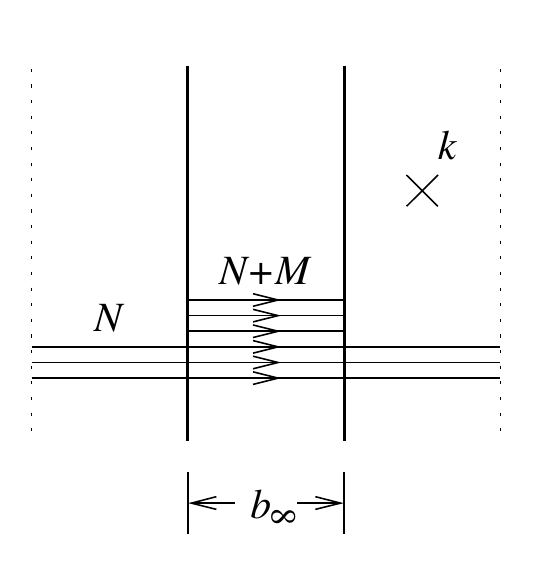} \\
(a) & (b)\\
\end{tabular}}
\caption{\label{figa} (a) Quiver diagram of $U(N+M)\times U(N)$ with $k$ flavors. (b) Brane construction for the theory studied in AHHO and CHH.  We have $M$ fractional D3 branes suspended between two NS5 branes in addition to $N$ integer branes.  The horizontal direction is periodically identified at the dotted lines.}
\end{figure}

The aim of this paper is to lay the foundation for such analysis.  For
a particular class of ${\cal N}=4$ theories in $2+1$ dimensions with
gauge and matter content illustrated in figure \ref{figa}.a, a gravity
dual in type IIA was constructed explicitly and was analyzed in some
detail by Aharony, Hashimoto, Hirano, and Ouyang (AHHO) in
\cite{Aharony:2009fc} and more recently by Cottrell, Hanson, and
Hashimoto (CHH) in \cite{Cottrell:2015jra}. In figure \ref{figa}.b, we
illustrate the brane construction for these models in type IIB string
theory. In the brane construction, one first constructs a defect field
theory in 3+1 dimensions on $R^{1,2} \times S^1$. One recovers the
theory in 2+1 dimensions by taking the radius of $S^1$ to zero keeping
the gauge coupling in 2+1 dimensions fixed.

These theories are formulated as $U(N+M) \times U(N)$ gauge field
theories with $k$ flavors charged under $U(N)$ and two bifundamental
matter fields in the ultra-violet. Such a system will then undergo a
renormalization group flow. The structure in the IR will depend on the
gauge group and the matter content. In \cite{Gaiotto:2008ak} Gaiotto
and Witten classified the IR dynamics of broad class of ${\cal N}=4$
theories in 2+1 dimensions into categories ``good,'' ``ugly,'' and
``bad.'' The ``good'' theories flows in the IR to interacting
superconformal fixed point, whreas the ``bad'' theories flows to a
trivial fixed point. For the class of models illustrated in figure
\ref{figa}.a, the condition for the theory to be ``good'' is
\be
\label{ftsusy}
0 \le -\min(M,0) \le N
\ee
and
\be
0 < -2 M < k \ . \label{gwgood}
\ee

The first condition (\ref{ftsusy}) is simply the requiment that the
rank of each gauge group in $U(N+M) \times U(N)$ is positive. 
The second is the condition for the gauge coupling not to diverge in 
the RG flow.

It would be interesting and desirable to identify the manifestation of
conditions (\ref{ftsusy}) and (\ref{gwgood}) in the gravity dual.
This very issue was studied recently in \cite{Cottrell:2015jra}, which
provided the following general picture. A supergravity ansatz can be set up, and solved, for any choice of $N$, $M$, and $k$ satisfying 
\be N - {M^2 \over k} \ge 0 \ . \label{sugbnd} \ee
Violating this bound will give rise to a repulson singularities. For
the set of repulson free solutions satisfying (\ref{sugbnd}), one can
explore the possibility of various brane probes becoming
tensionless. If that happens, we say that ``the background suffers
from enhancon effects \cite{Johnson:1999qt}.'' The analysis of
\cite{Cottrell:2015jra} revealed that the condition for the enhancon
not to appear is precisely (\ref{gwgood}).

\begin{figure}
\begin{center}
\includegraphics[scale = .8]{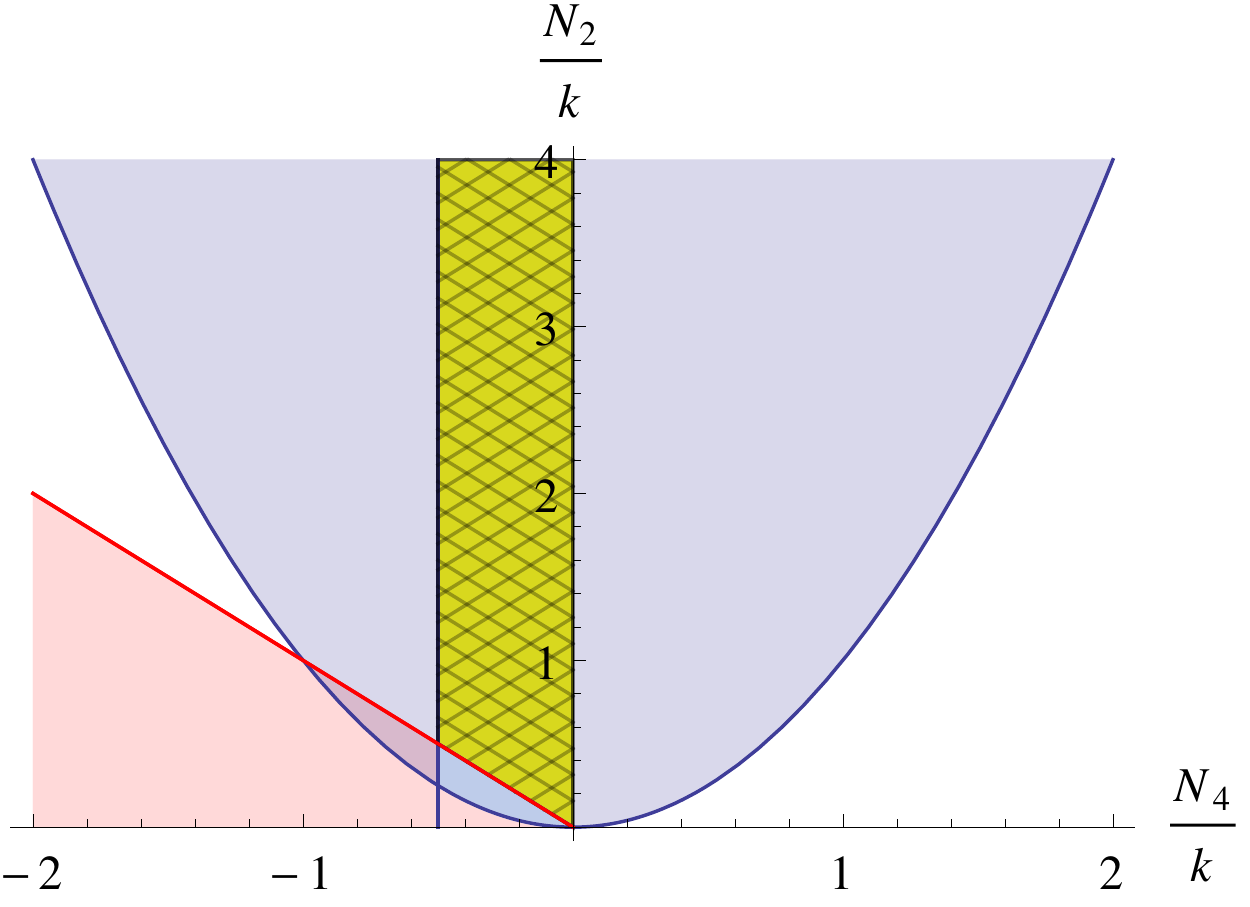}
\caption{One may partition the set of field theories according to various criteria.  Here, the parabola indicates theories with positive brane charge, the yellow cross-hatched strip labels the ``good'' theories and the red wedge region is the parts excluded by field theory considerations.}
\label{nodots}
\end{center}
\end{figure}

The remaining issue, then, is how to properly understand the apparent
discrepancy between (\ref{ftsusy}) and (\ref{sugbnd}).  A useful way
to highlight this issue is to work in the scaling limit
\be k \rightarrow \infty, \qquad {N \over k} = \mbox{fixed}, \qquad {M \over k} = \mbox{fixed.}  \label{kscale} \ee
This is somewhat like working in the 't Hooft limit for these
models. The features expected from the gauge theory perspective and
the gravity perspective for various choice of $N$ and $M$ is
illustrated in figure \ref{nodots}.

Before proceeding, let us note that scaling $N$ and $M$ like $k$ implies
\be N \ll k^5 \ee
and as such, we must use the type IIA description over the M-theory
description. The curvature radius in type IIA description then scales
as $R^2 \sim \alpha' \sqrt{N/k}$. These were the scaling found in
\cite{Aharony:2008ug} but they are applicable for our setup as well.

The proper understanding of the issue of (\ref{ftsusy})
v.s. (\ref{sugbnd}) which emerges is as follows. For the set of values
outside (\ref{gwgood}), the supergravity solution contains an
enhancon. As such, the issue of whether there is or isn't a repulson
at (\ref{sugbnd}) is meaningless. What we are saying is that inferring
(\ref{ftsusy}) on the gravity side outside (\ref{gwgood}) requires
{\it resolving} the dynamics of the enhancon. On the other hand, for
the models inside the range (\ref{gwgood}), the values of $N/k$ and
$M/k$ all take values of order one when either (\ref{ftsusy}) or
(\ref{gwgood}) are saturated.  This then suggests that (\ref{sugbnd})
receives $\alpha'$ corrections which when properly resummed reproduces
(\ref{ftsusy}). In other words, the apparent discrepancy between
(\ref{ftsusy}) and (\ref{sugbnd}) can be attributed entirely to
$\alpha'$ correction at least in the scaling regime (\ref{kscale}).

Now that we have a better understanding of the relationship between
(\ref{ftsusy}) and (\ref{sugbnd}), there are a number of interesting
directions one can explore. For the ``good'' theories satisfying
(\ref{gwgood}), one can use localization techniques
\cite{Nishioka:2011dq,Assel:2012cp} to compute the free energy
exactly. It would be interesting to recover (\ref{sugbnd}) as the
leading large $N/k$ approximation as well as subleading terms and
compare the first few corrections to curvature corrections on the
gravity side.

For theories not in the range (\ref{gwgood}), the task of resolving
the enhancon seems quite daunting. On the other hand, we know from
previous studies on related systems in 3+1 dimensions
\cite{Benini:2008ir} that these enhancons are closely associated with
the locus of enhanced gauge symmetry on the Coulomb branch as well as
being the baryonic root, a point on the Coulomb branch from which the
Higgs branch eminates \cite{Argyres:1996eh}.  More detailed analysis
of string correction in gauge gravity correspondence of ${\cal N}=2$
systems in 3+1 dimensions were carried out in
\cite{Cremonesi:2009hq,Conde:2013wpa}. It would be very interesting to
understand how the version of this story in 2+1 dimensions is
manifested on the gravity side of the gauge/gravity correspondence.

One way in which one might imagine approaching the baryonic root is to
consider turning on FI parameters. In gauge theory FI parameters 
generically smooth out the origin of Higgs branch and lift the Coulomb
branch. One can then approach the baryonic root by studying the limit
in which FI parameter is taken to zero. 

\begin{figure}[t]
\centerline{\includegraphics{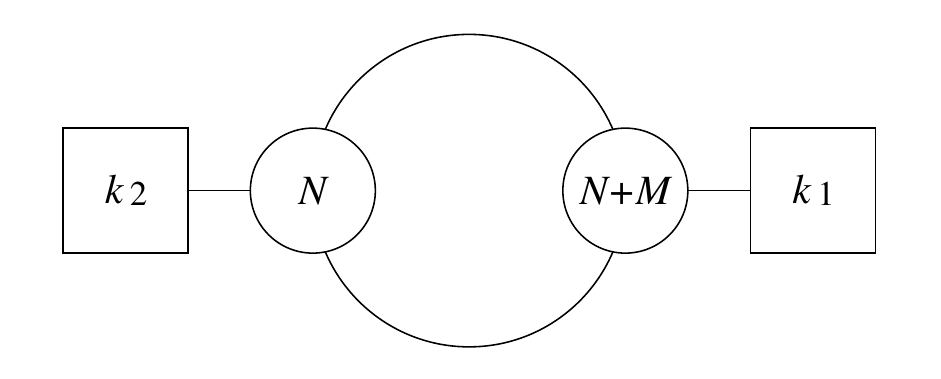}}
\caption{Quiver diagram for a theory with flavors charged under both gauge groups. \label{disflavors}}
\centerline{\includegraphics{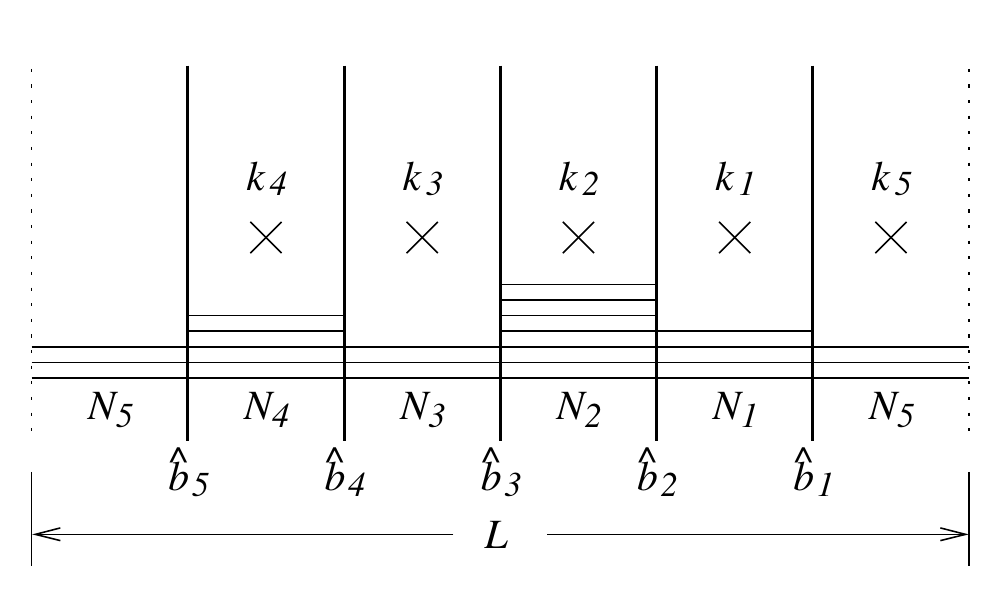}}
\caption{General circular quiver. \label{hwdiagram}}
\end{figure}

In order to fully explore the relationship between the field theory
and the gravity formulation of these ${\cal N}=4$ systems, it would
also be useful to have access to more general construction than the
class of models covered in figure \ref{figa}.a. One obvious
generalization is to allow matter to be charged under $U(N+M)$ as well
as $U(N)$, i.e.\  to consider a quiver of the type illustrated in
figure \ref{disflavors}.

It is not too difficult, it turns out, to generalize the construction
of the gravity solution to the one corresponding to the brane
construction illustrated in figure \ref{hwdiagram} with generic FI and
mass parameters turned on.

The goal of the remainder of this note is to describe such a
construction. As we will describe in detail below, considering generic
FI and quark mass naturally leads to the generalization of the ansatz
to include the possibility of adding matter charged under different
components of the product gauge group.  One can then consider the
limit of vanishing FI and mass parameters and obtain a candidate
gravity dual for the field theory with vanishing FI and mass
parameters at least for the particular point on the Coulomb branch
that one reaches in this limiting procedure. 

Another useful byproduct of considering generic FI and mass parameters
is the fact that the orbifolds which appear in the gravity dual are
completely resolved. This makes the analysis of charge quantization
much more straightforward and provides an independent derivation of
seemingly exotic charge and flux quantization relations outlined in
\cite{Aharony:2009fc}.

An important lesson we draw from these analyses is the fact that
$\alpha'$ corrections play a critical role in characterizing the
behavior of these systems close to the threshold of saturating the
conditions necessary for unbroken supersymmetry. It seems likely that
this is a generic fact about gauge gravity duality and implies that
understanding full string dynamics is required in order to study
phenomena such as dynamical supersymmetry breaking in gauge/gravity
duality \cite{Giecold:2013pza,Cottrell:2013asa}. It would be very
interesting to distill this issue and identify a tractable string
theory model which captures the dynamics of supersymmetric field
theories near the threshold of dynamical supersymmetry breaking
holographically, possibly along the lines of \cite{Michel:2014lva}.

The organization of this paper is as follows.  In Section \ref{review}
we review the basic setup and results of AHHO and CHH.  In Section
\ref{braneconstruction} we describe the general brane construction of
the theories we are interested in as well as the mapping to field
theory.  Next, in Section \ref{mtheory}, we show how to obtain sugra
solutions corresponding to these brane diagrams in M-Theory.  In
Section \ref{IIA} we also describe the IIA reduction and verify
certain aspects of the gauge/gravity duality using a probe analysis.
Finally, we offer our conclusions.

Some aspect of our supergravity construction, particularly the
enumeration of fluxes through compact cycles of the ALE background
geometry, can be found also in the previous work of \cite{Dey:2011pt}.
In our treatment, we provide additional consideration of fluxes
threaded by the non-compact cycles of the ALF which is related
parametrically to the coupling constants of the 2+1 dimensional gauge
theory in the UV. We also ellaborate on the quantization of fluxes and
charges from the IIA and M-theory perspective, and highlight our
expectations on correction due to  $\alpha'$ effects.

\section{Review of AHHO and CHH}
\label{review}

In \cite{Aharony:2009fc}, the authors considered a class of theories
consisting of ${\cal N}=4$ SYM in 2+1 dimensions with gauge group
$U(N+M)\times U(N)$ and $k$ fundamental hypermultiplets charged under
$U(N)$.  They are represented by a circular quiver of the form
illustrated in figure \ref{figa}.a. Such a model can be constructed
from the type IIB brane configuration illustrated in figure
\ref{figa}.b. The construction involves 2 NS5-branes and $k$
D5-branes, $N$ ``integer'' D3-branes winding all the way around the
$S^1$ of period $L$, and $M$ ``fractional'' D3-branes suspended
between the two NS5-branes separated by the distance $b_\infty L$. In
the $\alpha' \rightarrow 0$ zero slope limit, most of the string
states decouple and we obtain a 3+1 dimensional defect theory on
$R^{1,2} \times S^1$. In the limit that $L$ goes to zero while keeping
the gauge coupling in 2+1 dimensions fixed, momentum modes along the
$S^1$ decouples and we obtain a theory in 2+1 dimensions.

The gravity dual is constructed by T-dualizing along $S^1$ which maps
the 2 NS5-branes to $TN_{2}$ (which approaches the
$\mathbb{C}^2/\mathbb{Z}_2$ ALE geometry in the $L\rightarrow 0$
limit), D5-branes to D6-branes, integer D3-branes to D2-branes, and
fractional D3-branes to fractional D2-branes, which are D4-branes
wrapping the collapsed 2-cycle at the tip of the $C^2/\mathbb{Z}_2$
ALE.  

An important ingredient in understanding the gravity dual is the fact
that there are at least three notions of charge which become distinct
in the presence of fluxes. This issue was first emphasized by Marolf
\cite{Marolf:2000cb} who clarified the difference between brane, Page,
and Maxwell charges. Brane charges are localized and gauge invariant
but not quantized or conserved. Page charges are localized, quantized,
and conserved but not gauge invariant. Maxwell charges are conserved
and gauge invariant but not localized or quantized. In the context of
field theories of the type under consideration, this issue was
analyzed in detail in \cite{Aharony:2009fc}.  We will mostly follow
the conventions of \cite{Aharony:2009fc} and review these concepts as
needed. We also refer the reader to appendix B of
\cite{Aharony:2009fc} where a set of useful formulas are collected.
For the models depicted in figure \ref{figa}.a, $N$ is the D2 Page
charge and $M$ is the D4 Page charge.

One can then think of the IIA solution as a dimensional reduction of
M-theory on $R^{1,2} \times (C^2/\mathbb{Z}_2) \times TN_k$ to which
we add the back reaction of D2 and D4 branes sources. It is therefore
natural to consider an ansatz where $R^{1,2} \times (C^2/\mathbb{Z}_2)
\times TN_k$ gets warped as a result of fluxes sourced by the D2 and
the D4-branes.
 
The ansatz considered in  \cite{Aharony:2009fc} is
\beq ds^2 &=& H^{-2/3} (-dt^2+dx_1^2+dx_2^2) + H^{1/3} (ds_{ALE}^2+ds_{TN_k}^2), \\
G_4 &=& dC_3 =  dt \wedge dx_1 \wedge dx_2 \wedge d H^{-1} + G_4^{SD}, \label{g4}\\
G_4^{SD} &=& d(l V \omega_2 \wedge \sigma_3 + 2 \alpha \omega_2 \wedge d \psi)\ .  \label{ahhoansatz} \eeq
Our conventions are as follows:
\begin{itemize}
\item The 2-form $w_{2}$ is a self-dual 2-form on the collapsed cycle of the ALE normalized by:
\be
\int_{ALE} w_{2}\wedge w_{2} = \frac{1}{2}
\ee
\item The M-Theory circle is parameterized by $\psi$ living in the (unresolved) $TN_{k}$, whose metric is given by:
\be
ds^{2}_{TN_{k}} = V(r)^{-1}\left(dr^{2}+r^{2}\left(d\theta^{2}+\sin^{2}\theta d\phi^{2}\right)\right) + V(r)R_{11}^{2}k^{2}\left(d\psi-\frac{1}{2}\cos\theta\phi \right)^{2}
\ee
with
\be
V(r) \equiv \left(1+\frac{k R_{11}}{2r}\right)^{-1},\,\,\,\,\,\,\,\,\,\,\,\, R_{11} = l_{s}g_{s} \ . 
\ee
\item $\sigma_{3}$ is a one-form form living on $TN_{k}$  defined as\footnote{The subscript `$3$' on this form comes from its role as the Cartan-Weyl one-form under the identification $SU(2)= S^{3}$.} 
\beq
\frac{1}{2}\sigma_{3} = d\psi - \frac{1}{2} \cos(\theta)d\phi
\eeq
where $\psi = \psi + 2\pi/k$.  
\item We may also introduce a parameter $Q_{2}^{brane}$ in the sugra
  equation for the warp factor
\be 0 =  \left(\nabla^2_y+ \nabla^2_{TN} \right) H + {l^2 V^4\over 2  r^4}  \delta^4(\vec y)
+ (2 \pi l_p)^6  Q^{brane}_2 \delta^4(\vec y) \delta^4(\vec r)
\label{harm} \ . \ee
\end{itemize}

Note that despite the fact that the parameter $\alpha$ in equation
(\ref{ahhoansatz}) is the coefficient of a total derivative term and
would hence seem to be trivial, it turn out to be integrally quantized
and related to the D4 Page charge via:
\be
\label{alpha}
2\pi \alpha = (2\pi l_{s})^{3} g_{s} M \ . 
\ee
Also, $l$ is determined in terms of the Page charge, $M$, and $b_{\infty}$ via:
\be
l = -(2\pi l_{s})^{3}g_{s}\left(M+\frac{k b_{\infty}}{2}\right) \ . 
\ee
The only remaining ingredient is the brane charge $Q_{2}^{brane}$.  This was found to be:
\be
Q_{2}^{brane} = N - \frac{M^{2}}{k} =\frac{R_{AdS_{4}}^{6}}{64 k \pi^{2} l_{p}^{6}} \ . \label{Q2b} 
\ee
When $Q_{2}^{brane} <0$, the warp factor (as determined by equation
(\ref{harm})) becomes negative at sufficiently small radii giving rise
to a repulson singularity.  In general, for theories with $8$ real
susy generators, we expect that such singularities are always masked
within an enhancon sphere at an even larger radius.  This is indeed
what was found in \cite{Cottrell:2015jra}.  The enhancon radius is
defined by the appearence of tensionless probes, which is equivalent
to the divergence of the effective gauge coupling.

To study this issue, we may insert probe D4 branes wrapping the
collapsed cycle of the $ALE$.  One may label these probes by the D4
charge and dissolved D2 charge.  One finds that these experience a
Taub-NUT moduli space with NUT charges of $nk\mp 2M$, where $n=\mp \#
D2$ and the upper/lower sign is taken for positive/negative D4
charge.  Using the appropriate probes, one finds that the  strength of the gauge couplings can be parameterized by a dimensionless parameter
\beq
\label{rgflow}
\frac{1}{g_{eff1}^{2}} &=& b_{\infty} -\frac{g_{YM}^{2} 2M}{4\pi \Phi} \\
\frac{1}{g_{eff2}^{2}} &=& (1-b_{\infty}) +\frac{g_{YM}^{2}(k+2M)}{4\pi \Phi}
\label{rgflow2}
\eeq
where 
\be g_{eff1,2} = {g_{YM1,2}^2 \over g_{YM}^2} \ee
for gauge groups $U(N)$ and $U(N+M)$, respectively, with
\be {1 \over g_{YM}^{2}} = {1 \over g_{YM1 }^{2}}+{1 \over g_{YM2}^{2}}= g_{s}/l_{s} \ee
and 
\be \Phi = {1 \over 2 \pi \alpha'} r \ee
is the scale of renormalization being probed by the brane.

From the form of the probe action, it is straightforward to infer if
and when an enhancon will arise from a probe whose (D2,D4) charge is
$(n,\pm 1)$. This happens whenver $1/g_{eff1}^2$ or $1/g_{eff2}^2$
vanishes. It is clear from (\ref{rgflow}) and (\ref{rgflow2}) that
this will not happen provided (\ref{gwgood}) is satisfied. This is how
one sees (\ref{gwgood}) arising on the supergravity side as was
described in \cite{Cottrell:2015jra}.

One natural generalization to this class of models is to include
flavors charged under $U(N+M)$ as well as $U(N)$, as illusterated in
figure \ref{disflavors}.  In \cite{Aharony:2009fc}, it was conjectured
based on consideration of brane shuffling analysis that the gravity
dual for these models should correspond to shifting the Page
charge\footnote{See equation (2.82) of \cite{Aharony:2009fc}}
\beq
Q_{2}^{Page} &=& N +\frac{k_{1}}{4} \label{pageanomaly} \\
Q_{4}^{Page} &=& M - \frac{k_{1}}{2}\ .  \label{pageanomaly2}
\eeq
However, a formal derivation of this relation from a purely
supergravity consideration was not provided. We will show in the
following sections that the conjecture of \cite{Aharony:2009fc} is
indeed correct.

\section{Brane Construction}
\label{braneconstruction}

The class of theories that we are interested in may be represented by Hanany-Witten brane constructions consisting of NS5, D5, and D3 branes with the following orientation:

\be
\begin{tabular}{c||cccccccccc}
       & 0& 1  & 2& 3& 4& 5& 6& 7& 8& 9 \\
       \hline
NS5 & $\circ$ &  $\circ$ & $\circ$ &  $\circ$ &  $\circ$  &$\circ$   &   &   &   &     \\
D5 & $\circ$ & $\circ$  &$\circ$& & &    &  &$\circ$ &$\circ$ &$\circ$     \\
D3 & $\circ$& $\circ$& $\circ$& &  &  & $\circ$ &  &   &
\end{tabular} \label{orientation}
\ee

Arranging these elements as in figure \ref{hwdiagram}, we may
construct an $\mathcal{N}=4$ theory in $2+1$ dimensions with gauge
group of the form $\prod_{i=1}^{\hat{k}}U(N_{i})$, bi-fundamental
hypermultiplets charged under neighboring gauge groups, and $k_{i}$
hypermultiplets charged under $U(N_{i})$.  The coordinates $3,4,5$
labeling the transverse D5 position will be denoted by a set of $k$
3-vectors, $\vec{x}_{j}$.  Likewise, the transverse positions of the
$\hat{k}$ NS5 branes will be denoted by $\hat{k}$ vectors,
$\vec{y}_{i}$, labeling the $7,8,9$ coordinates.

In drawing figure \ref{hwdiagram} we have chosen the $U(N_{\hat k})$
interval as the one to ``cut'' open to present the circular quiver in a
linear form.  This allows us to define the concepts of ``left'' and
``right'' in a circular quiver.  Opening the quiver at a node other than
$U(N_{\hat k})$ will permute the notation but not otherwise affect the
physics in any way.

In this diagram, $\hat{b}_{i}$ labels the position of the $i$-th NS5
brane (denoted NS5$_{i}$) before brane bending in the sense of
\cite{Witten:1997sc} is considered.  In other words, $\hat{b}_{i}$ is
the limit of the $x^{6}$ position of NS5$_{i}$ as
$|\vec{x}|\rightarrow \infty$.  We will choose the indexing of the
branes such the the sequence
$(\hat{b}_1,\hat{b}_{2}...,\hat{b}_{\hat{k}})$ is non-increasing as
in figure \ref{hwdiagram} and the $x^{6}$ origin will be defined so
that $\hat{b}_{\hat{k}} = 0$. This then gives rise to $\hat k -1$ continuous
parameters. The choice of ordering is made so as to gaurantee that the
asymptotic gauge couplings will be positive.  The UV couplings may be
read off from the HW diagram as:
\be \frac{1}{2 g_{YMi}^{2}} =
\frac{L\left(\hat{b}_{i}-\hat{b}_{i+1}\right)}{g_{s}^{IIB}} \ . 
\label{gvb}
\ee

We may also assign an asymptotic position, $b^{j}(\infty)$, to each D5
brane. This would increase the number of continous paramters from
$\hat k-1$ to $\hat k + k - 1$. The position of $b^k$ relative to
$\hat b_{\hat{k}}$ is physcally meaningful. We will see below that as
one flows from defect field theory in 3+1 dimensions to the gauge
theory in 2+1 dimensions, these $k$ degrees of freedom parameterized
by $b^j$ decouples.

\begin{figure}
\centerline{\includegraphics{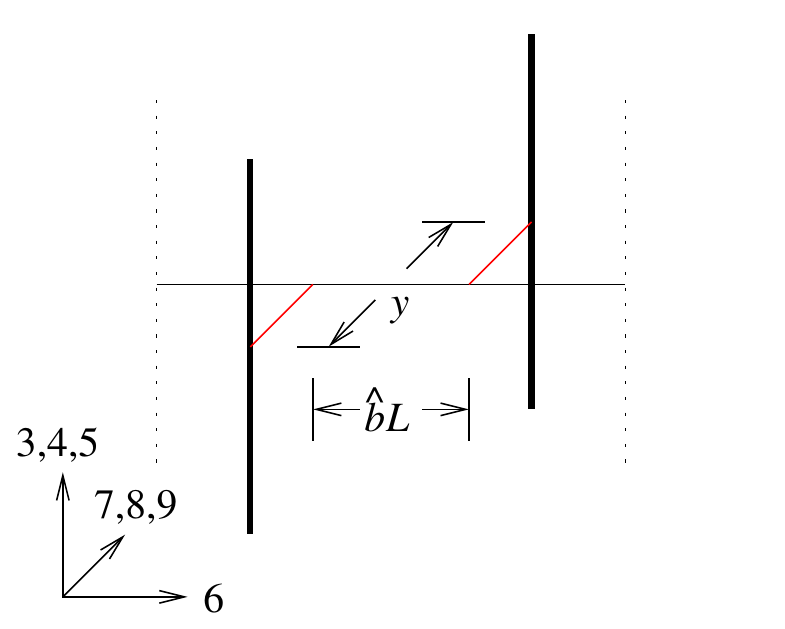}}
\caption{Determination of field theory FI parameter using BPS properties and s-duality.  The black dots represent NS5 branes that are displaced in their transverse dimensions by a distance $y$.  The periodicity of $x_{6}$ is taken to be $L$.  The red lines represent D1 strings breaking on the D3 and their BPS mass determines the FI term.}
\label{FIterm}
\end{figure}

Resolving the ALE and the Taub-NUT corresponds to shifting the NS5 and
the D5-branes in the transverse space, respectively.  These, in turn,
correspond to FI and mass parameters in the field theory,
respectively.  The relationship between the D5 position and the field
theory mass parameter is easy to determine since the mass parameter is
simply the mass of a fundamental string stretching between the D5 and
a D3 stack.  The scaling of the FI parameters may similarly be fixed
in terms of the mass of a D-string as is illustrated in figure
\ref{FIterm}.  We find:
\beq
\label{mass}
\vec{m}_{j} &=& \frac{\vec{x}_{j}}{2\pi \alpha'} \\
\label{FI}
\vec{\xi}_{i} &=& \frac{\vec{y}_{i}}{2\pi g_{s}^{IIB} \alpha'} \ . 
\eeq

\begin{figure}
\centerline{\includegraphics{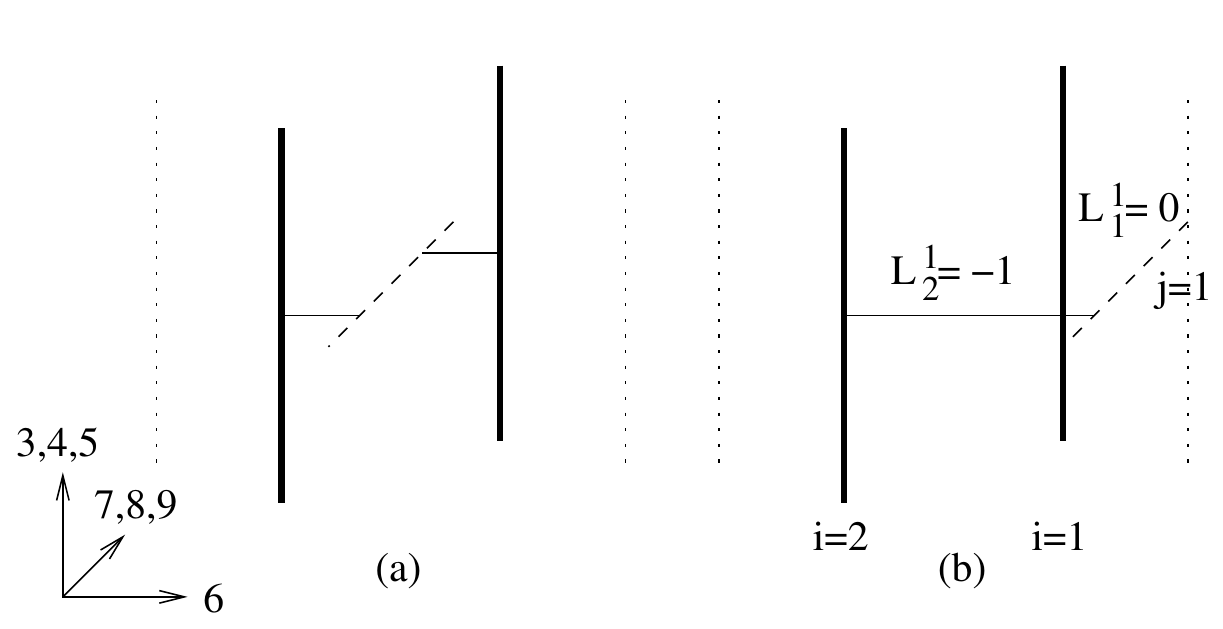}}
\caption{\label{conv2} Hanany-Witten diagram (a) for a configuration allowing FI deformation without breaking any supersymmeries, and  (b) the same configuration with the D5-branes pushed to right such that the data $L_i^j$ is manifest.}
\end{figure}

For generic $\vec{x}_{j}$ and $\vec y_i$, it is no longer possible for
a D3 to end on a pair of D5 branes or NS5 branes while preserving
supersymmetry.  There are, however, supersymmetric configurations
corresponding to D3 segments extending between a D5/NS5 pair as
in figure \ref{conv2}.  The brane configuration is thus specified by
giving the number of D3 segments between each NS5$_{i}$ and D5$_{j}$
pair for the given choice of cut.  For the sake of systematically
displaying this data, it is more convenient to move all the D5
branes to the right using Hanay-Witten transitions
\cite{Hanany:1996ie}. We will end up with a diagram such as figure
\ref{conv2}.  In figure \ref{conv2}.b we illustrate a possible pattern
of linkings corresponding to a resolved version of figure
\ref{conv2}.a.  The parameters $L_{i}^{j} \in \mathbb{Z}$ characterize
the allowed configurations and correspond to the number of D3
segments extending from the $i$-th NS5 to the $j$-th D5.

\begin{figure}
\centerline{\includegraphics[width=\hsize]{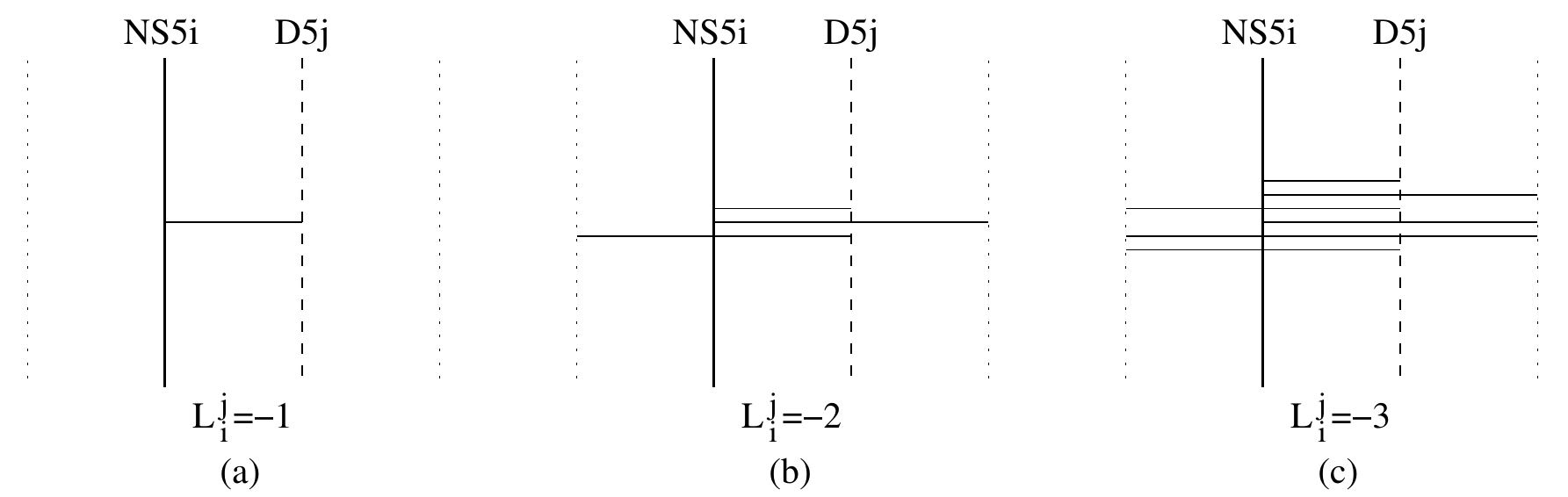}}
\caption{\label{Lij} Multiple wrappings are allowed on a circular quiver.}
\end{figure}

Note that on a linear quiver the $s$-rule implies that only a single
D3 segment may connect a given NS5/D5 pair, hence implying $L_i^j$
taking values\footnote{The orientation is chosen such that negative
  $L_{i}^{j}$ corresponds to a D3 segment ending on the left of the
  D5, while positive $L_{i}^{j}$ corresponds to a D3 extending from
  the right.  }, 0, 1, or $-1$.  This restriction is weaker on a
cricular quiver since we can allow the D3 segment to wrap multiple
times as in figure \ref{Lij}.  This is equivalent to satisfying the
$s$-rule in the covering space.  For any given value of $L_{i}^{j}$,
we will always take the brane configuration to be the one with the
minimal number of wrappings consistent with the $s$-rule. For example,
in figure, \ref{Lij}, we must wrap the fractional brane, $0$, $1$, and
$3$ times for $L_{i}^{j} = -1,-2,-3$, respectivley.  We could consider
introducing more than the minimal number of wrappings at each step;
for example, we could have introduced $0$, $2$, and $5$ wrappings
instead for the same sequence of $L_{i}^{j}$'s.  However, the freedom
to do this is already encompassed by the freedom to add $N_2^{free}$
integer branes.  The integer $N_2^{free}$ will turn out to be related
to the number of M2-branes one needs to add to the resolved $ALE
\times TN$ space. For now, we will analyze the configuration where
$N_2^{free}$ is set to zero, corresponding to a configuration where
all of the Coulomb-branch is lifted. It will be straightforward to add
these extra $N_2^{free}$ branes at a later stage.

In figure \ref{Lij}, for instance, we see that for $L_{i}^{j} =
-1,-2,-3....$ we need to introduce at least $|L_{i}^{j}|-1$ wrappings
at the last step.  Thus, the total number of wrappings associated with
$L_{i}^{j}$ that we {\it{must}} add is equal to:
\beq
\# \text{wrappings}&=& 0+1+2+....+(|L_{i}^{j}|-1) \cr
&=&\frac{|L_{i}^{j}|(|L_{i}^{j}|-1)}{2} \label{wrappings} \\
&=& \frac{L_{i}^{j}\left(L_{i}^{j}+1\right)}{2} \ .  \nonumber
\eeq
In the last step we used the fact that $L_{i}^{j} <0$ in the example
above.  If $L_{i}^{j}$ had been positive, similar logic would have
yielded a total number of extra wrappings equal to
$L_{i}^{j}(L_{i}^{j}-1)/2$.

A useful intermediate concept is the linking numbers
\cite{Hanany:1996ie}.  These are essentially monopole charges on the
world-volumes of the 5-branes and their utitlity lies in the fact that
they are invariant under Hanay-Witten brane maneuvers.  Following,
\cite{Assel:2011xz}, we take the linking numbers to be defined as
\beq
\label{linkdef}
l^{j} &=& (\text{net D3's ending on left}) + (\text{NS5's to the right}) \\ \nonumber
\hat{l}_{i} &=& (\text{net D3's ending on right}) + (\text{D5's to the left}) \ . 
\eeq
An inspection of the diagram shows that the linking numbers and the
$L_{i}^{j}$ are related by:
\be
\label{linkLl}
l^{j} = -\sum_{i=1}^{\hat{k}} L_{i}^{j}, \qquad 
\hat{l}_{i} = -\sum_{j=1}^{k} L_{i}^{j}  \ . 
\ee

It should be clear that this map from the brane data $L_{i}^{j}$ to
linking numbers is many to one. Indeed, at this stage, $L_{i}^{j}$ is
an unconstrained $k \times \hat{k}$ array while $\hat{l}_{i}$ and
$l^{j}$ are merely $\hat{k}+k$ variables with one constraint;
$\sum_{i}\hat{l}_{i}=\sum_{j}l^{j}$.  We therefore have
$k\hat{k}-k-\hat{k}+1$ extra pieces of information.  This redundancy
exist in the map from $L_{i}^{j}$ to discrete field theory data,
$(N_{i},k_{i})$, as well.  Some of the extra information corresponds
to a choice of vacuua within the specified theory.

However, not all configurations parameterized by distinct values of
$L_{i}^{j}$ are physically distinct when classifying the corresponding
field theory vacua in $2+1$ dimensions.  If we rotate the $j$-th D5
around the circle $p^{j}$ times then the linking number of this particular
D5 will change by $-\hat{k}p^{j}$ and the linking number of
each NS5 will change by $-p^{j}$.  It is straightforward to check
that the full set of such transformations allows one the freedom to
shift $L_{i}^{j}$ (and hence $\hat{l}_{i},\,l^{j}$) as
\beq
\label{lgauge}
L_{i}^{j} &\rightarrow& L_{i}^{j}+ p^{j} \\ \nonumber
l^{j} &\rightarrow & l^j - \hat{k} p^j  \\ 
\hat{l}_i & \rightarrow& \hat{l}_i  - \sum_{j=1}^{k} p^j \nonumber
\eeq
where $p^{j}$ is an integer.  We remark that the convention
illustrated in figure \ref{Lij} for defining the brane configuration
in terms of $L_{i}^{j}$ is invariant under this operation.  This
suggest that we may shift to
\be
\label{branegauge}
0 \le l^{j} \le \hat{k}-1
\ee  
and obtain all physically distinct configurations.  This is possible
because the position of the D5 is irrelevant and has decoupled in
flowing to the $2+1$ dimensional field theory.  The condition
(\ref{branegauge}) may be implemented by using (\ref{lgauge}) to
replace the original $L_{i}^{j}$ by
\be
\label{transformed}
L_{i}^{j}\rightarrow L_{i}^{j} - \lceil \frac{1}{\hat{k}} \sum_{p=1}^{\hat{k}} L_{p}^{j}\rceil
\ee
where $\lceil \cdot \rceil$ is the ceiling function.  In the following
formulas we will assume that $L_{i}^{j}$ already satisfies
(\ref{branegauge}) and so suppress the transformation
(\ref{transformed}).

\begin{figure}
\begin{center}
\begin{tabular}{cc}
\includegraphics[scale=0.9]{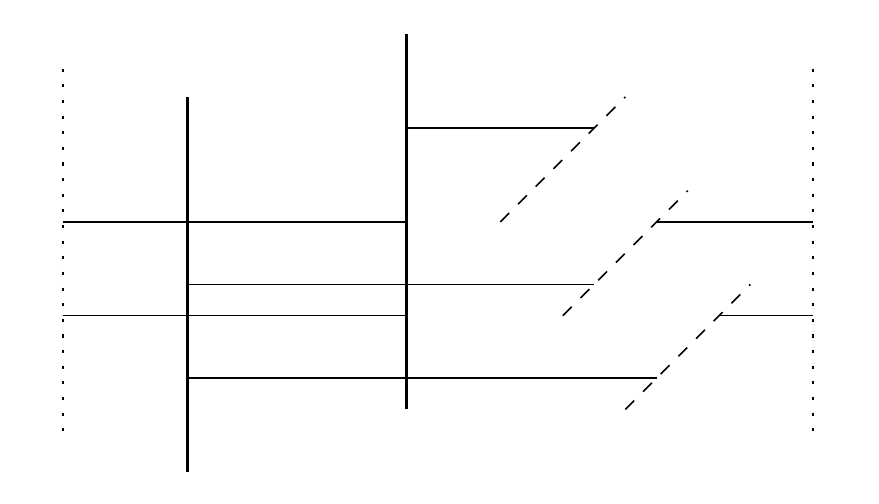} & \includegraphics[scale=0.9]{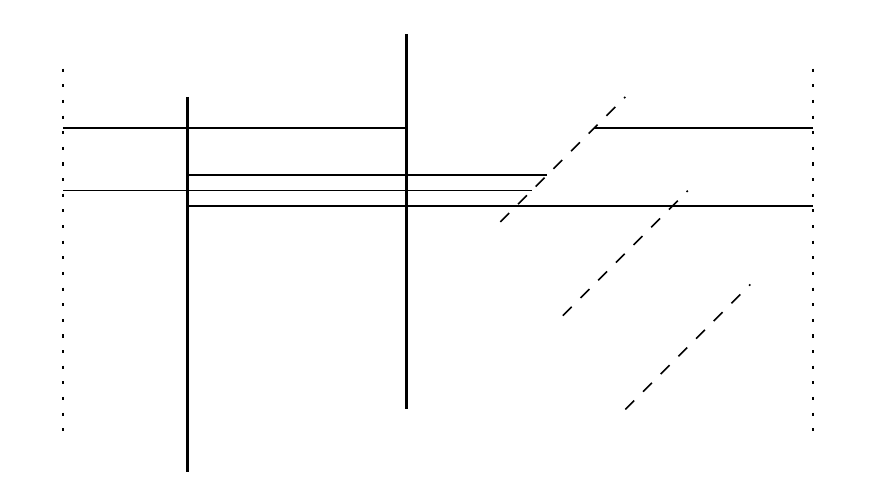} \\
$L_i^j = \left(\begin{array}{ccc}
-1  & 1 & 1 \\
0 & -1 & -1
\end{array}\right)$ &  
$L_i^j = \left(\begin{array}{ccc}
1  & 0 & 0 \\
2 & 0 & 0
\end{array}\right)$ 
\end{tabular}
\end{center}
\caption{Different choices for $L_{i}^{j}$ may give rise to the same field theory in the orbifold limit.
\label{many2one}}
\end{figure}

%

Now that we have enumerated the brane configurations (for
$N_2^{free}=0$) in terms of $L_i^j$ and $\hat b_i$, it would be useful
to understand how these data are related to the data $N_i$ and $k_i$
which appear in figure \ref{hwdiagram}. It is in terms of the latter
parameterization that the gauge and matter content of the field theory
is easy to read off.

This can be acheived by basically inverting the process illustrated in
figure \ref{conv2} by moving the D5-branes to the right until there
are no D3-branes ending on them. The condition (\ref{branegauge}) will
guarantee that this procedure will terminate before the D5-branes
circumnavigate the periodic $x_6$ direction.  One can easily infer
that this procedure leads to the relation
\beq
\label{keq}
k_{i} &=& \#\{j|l_{j} = i \mod \hat{k}\} \\
\label{nieq}
\hat{l}_{i} &=& N_{i-1}-N_{i}+\#\{j | l_{j} \ge i\} \ . 
\eeq
The exercise was carried out in section 2.3 of
\cite{Gaiotto:2008ak}. Our (\ref{nieq}) is identical to (2.5) of
\cite{Gaiotto:2008ak}.

The relation (\ref{nieq}) can be used to partially solve for $N_i$'s
\be
\label{nianswer}
N_{i} = \sum_{p=1}^{i}\sum_{j=1}^{k} L_{p}^{j} + \sum_{p=1}^{i} \#\{j | l_{j} \ge p\}+N_{\hat k} \ . 
\ee
The relation (\ref{nieq}) does not contain the information needed to
fix $N_{\hat k}$, but that can also be extracted from the $L_i^j$ data.

The issue is closely related to the counting of wrappings discussed
previously in (\ref{wrappings}). It is easy to see the pattern by
working out few simple cases with small values of $L_i^j$ with some
care in considering the cases with positive and negative values of
$L_i^j$. Regardless, one can show that
\be N_{\hat k} = N_2^{free} + {1 \over 2} \sum_{i=1}^{\hat k}  \sum_{j=1}^{k} (L_i^j(L_i^j+1)) \label{n0answer}\ee
where for sake of completeness, we included the possibility to add
arbitrary non-negative integer $N_2^{free}$ corresponding to
additional integer branes. The smallest allowed value of $N_{\hat{k}}$ then
corresponds to setting $N_2^{free}$ to zero, and that
\be N_{\hat k}^{min} = {1 \over 2} \sum_{i=1}^{\hat k}  \sum_{j=1}^{ k} (L_i^j(L_i^j+1))\ee
which can also be seen to be strictly non-negative provided $L_i^j$'s
are integer valued. Since the choice of which of the $N_i$'s to
identify as $N_{\hat{k}}$ was arbitrary, it also follows that all of the
$N_i$'s must be positive definite.

\subsection{Summary of Brane Construction}
\label{FTG}

Since quite a bit of technical issues were discussed in this section,
let us pause and summarize the main results. We enumerated the
distinct brane configruations consisting of $\hat k$ NS5-branes and
$k$ D5-branes in terms of the following data\\

\centerline{\begin{tabular}{|l|l|} \hline
$L_i^j$ & Number of D3-branes stretching between NS5$_i$ and D5$^j$ \\ \hline
$N_2^{free}$ & Number of additional integer D3-branes \\ \hline
$\vec x_j$ & Position of D5$^j$ in $x_{3,4,5}$ coordinates \\ \hline
$\vec y_i$ & Position of NS5$_i$ in $x_{7,8,9}$ coordinates \\ \hline
$\hat b_i$ & Position of NS5$_i$ the periodic $x_6$ coordinate \\ \hline
\end{tabular}}

\noindent with the following additional comments.
\begin{itemize}

\item $i$ takes values in the range $1 \le i \le \hat k$ and $j$ takes
  values in the range $1 \le j \le k$.

\item $L_i^j$ consists of $k \times \hat k$ integers subject to
  constraint (\ref{branegauge}) modulo permutation of $j$.

\item $N_2^{free}$ is a non-negative integer.

\item $\hat b_i$ gives rise to $\hat k-1$ continious parameters taking
  values in the range $0 = \hat{b}_{k} \le \hat b_{\hat k-1} \le
  \ldots \le \hat b_2 \le \hat b_1\le 1$. $\hat b_i$ is
  dimensionless. The physical position in the $x_6$ coordinate is
  given by $x_{6i} = L \hat b_i$ where $L$ is the period of the $x_6$
  coordinate.

\end{itemize}

These data were then mapped to\\ 

\centerline{\begin{tabular}{|l|l|} \hline
$N_i$ & Rank of $U(N_i)$  gauge group in a circular quiver \\ \hline
$k_i$ & Number of fundamtals charged with respect to $U(N_i)$ \\ \hline
$\vec \xi_i$ & Fayet-Illiopolous parameter for $U(N_i)$. \\ \hline
$\vec m_j$ & Mass of the $j$-th fundamental charged with respect to $U(N_i)$ for $i=(l_j \mod \hat k)$\\ \hline
$g_{YMi}^2$ & Gauge coupling of $U(N_i)$  \\ \hline
\end{tabular}}

The map relating these to sets of data were presented in equations
(\ref{gvb}), (\ref{mass}) ,(\ref{FI}), (\ref{keq}), (\ref{nianswer}),
and (\ref{n0answer}). For the convenience of the reader, they are also
collected below.
\beq
\frac{1}{2 g_{YMi}^{2}} &=& \frac{L\left(\hat{b}_{i}-\hat{b}_{i+1}\right)}{g_{s}^{IIB}} \label{pimap1}\\ 
\vec{m}_{j} &=& \frac{\vec{x}_{j}}{2\pi \alpha'} \\
\vec{\xi}_{i} &=& \frac{\vec{y}_{i}}{2\pi g_{s}^{IIB} \alpha'} \\ 
k_{i} &=& \#\{j|l_{j} = i \mod \hat{k}\} \\ 
N_{i} &=& \sum_{p=1}^{i}\sum_{j=1}^{k} L_{p}^{j} + \sum_{p=1}^{i} \#\{j | l^{j} \ge p\}+N_{\hat k} \\
N_{\hat k} &=& N_{2}^{free}+\frac{1}{2} \sum_{i=1}^{\hat k}\sum_{j=1}^k \left((L_{i}^{j})^{2}+L_{i}^{j}\right) \ . \label{pimap2}
\eeq

The discrete data are contained in $L_i^j$ or the set $(N_i,k_i)$. As
we noted previously, the mapping is not one to one. This reflects the
fact that a field theory identified by $(N_i,k_i)$ might admit more
than one vacua. The $L_i^j$ provides a parameterization which is
distinct for each of these vacua.

It would be useful to develop some feel with regards to which values
of $(N_i,k_i)$ arise as corresponding to some $L_i^j$. This is
equivalent to enumerating the set of $(N_i,k_i)$ which admits a
supersymmetric vacuum for a generic value of the FI and the quark mass
parameters. Unfortunately, the relations
(\ref{pimap1})--(\ref{pimap2}) are a bit too cumbersome to convey that
intuition in general, but one can easily analyze this issue explicitly
for simple cases such as taking $\hat k =2$.  This is precisely the
case where the quiver diagram takes the form illustrated in figure
\ref{disflavors}.

\begin{figure}
\begin{center}
\includegraphics[scale = .8]{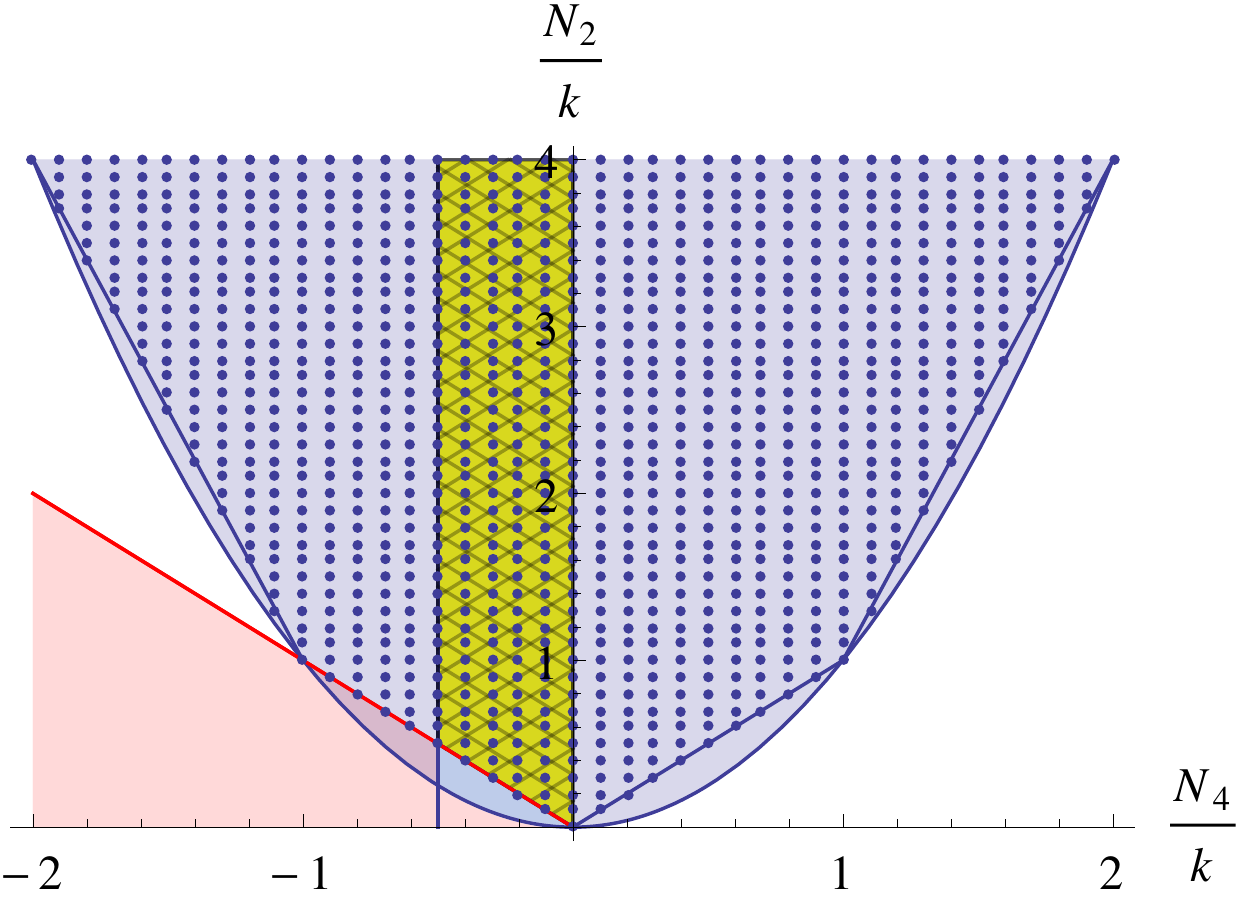}
\caption{Dots represent set of resolvable theories for $\hat{k}=2,\,k=10,\,k_{1}=0$.  The yellow region is the ``good'' region for which no enhancon appears.}
\label{akiplot}
\end{center}
\end{figure}

For this case, equations  (\ref{linkLl}), (\ref{branegauge}), and (\ref{keq}) simplify to 
\be
l^j = -L_{1}^{j}-L_{2}^{j} = 
\begin{cases} 
0 & \quad \text{for} \quad  1\le j\le k_{1} \\
1 & \quad \text{for} \quad k_{1} < j \le k \ . 
\end{cases} \label{lLL}
\ee
We may use this to eliminate $L_{2}^{j}$ in the formulas
(\ref{n0answer}) and (\ref{nianswer}) for $N_{1}$ and $N_{2}$. Let us
further restrict to the case where $k_{2} = k$, corresponding to the
quiver illustrated in figure \ref{figa}.a. Then, for $N_{2}=N$ and
$M=N_{1}-N_{2}$, we find
\beq
\label{rescriteria}
N&=& N_2^{free}+ \sum_{j=1}^{k} (L_{2}^{j})^{2} \\ 
M&=&-\sum_{j=1}^{k}L_{2}^{j}
\eeq
for the allowed values of $N_2^{free}$ and $L_i^j$. Figure
(\ref{akiplot}) illustrates the set of $N$ and $M$ which can arise
this way. As noted earlier, these correspond to the set of $(N,M)$ for
which a supersymmetric state exists for generic values of $\vec x_j$
and $\vec y_i$. For now, let us simply note that
\begin{itemize}
\item The set of $(N,M)$ closely resembles the parabola $Q_2^{brane}\ge 0$ for $Q_2^{brane}$ given in (\ref{sugbnd}) but misses some of the regions even in the large $k$ limit.
\item The set of $(N,M)$ is contained in but does not saturate the region defined by (\ref{ftsusy}).
\end{itemize}

\section{Gravity Dual Description}
\label{mtheory}

In this section, we construct the gravity duals of brane configuration
enumerated in the previous section. Just like in the examples
considered in \cite{Aharony:2009fc} and \cite{Cottrell:2015jra}, start
by applying T-duality along $x_6$ direction to map the system to IIA,
and lift to M-theory. The $\hat k$ NS5-branes and $k$ D5-branes in the
original IIB frame will map to a space whose structure is roughly that
of $R^{1,2} \times TN_{\hat k} \times TN_k$ where the subscript denote
the Taub-NUT charge. Since we are considering the setup where the
$\hat k$ FI parameters and $k$ mass parameters are allowed to take
generic values, we will treat these multi-centered Taub-NUT geometry
is generically resolved.

Adding integer and fractional branes then corresponds to adding
sources and letting the geometry warp. Just as was the case in earlier
work \cite{Aharony:2009fc,Cottrell:2015jra}, one can write an explicit ansatz for the fields in 11 dimensional supergravity of the form
\beq
\label{ansatz}
ds^{2}&=&H^{-2/3}\left(-dt^{2}+dx_{1}^{2}+dx_{2}^{2}\right)+H^{1/3}\left(ds^{2}_{TN_{\hat{k}}}+ds^{2}_{TN_{k}}\right) \\  \nonumber
G_{4} &=&-dH^{-1} dt\wedge dx_{1}\wedge dx_{2} + G_{4}^{SD} 
\eeq
and check that the BPS condition and subsequently the equation of
motion is satisfied, provided that $G_4^{SD}$ is a self-dual 4-form in
$TN_{\hat k} \times TN_k$, and the warp factor $H$ satisfies a Poisson
equation
\be
\label{warpfactor}
\nabla_{TN_{\hat k} \times TN_k}^{2}H = -\frac{1}{2} *_8 (G_{4}^{SD}\wedge G_{4}^{SD})
- \sum_{m=1}^{Q_{M2}} (2\pi l_{p})^{6}  \delta^{8}(\vec{z}-\vec{z}_m))
\ee   
also on $TN_{\hat k} \times TN_k$. Here, $\vec{z}_m$ denotes the position of an M2-brane in $TN_{\hat k} \times TN_k$.

Some properties of resolved Taub-NUT geometry will play an important role in the analysis below, so let us briefly recall the key facts.

A $k$-centered Taub-NUT is a four dimensional Euclidian geometry which can be viewed as an $S^1$ fibered over an $R^3$. It has a metric
\be ds^{2}_{TN_{k}}=V R_{11}^2\left(d\psi- \frac{1}{2} \mathbf{w}\right)^{2}+V^{-1} d\vec{x}\cdot d\vec{x}\, 
\ee
where $V(\vec x,\psi)$ is a scalar function
\be V^{-1}=1+\sum_{j=0}^{k-1}\frac{R_{11}}{2|\vec{x}-\vec{x}_{j}|}\,  \ee
and
\be \mathbf{w} = \vec{\mathbf{w}} \cdot d \vec x \ee
is a one form, and a vector $\vec{\mathbf{w}}$ is defined by 
\be 
\frac{2}{R_{11}}\vec{\nabla} V^{-1} = \vec \nabla \times  \vec{\mathbf{w}}  \ . 
\ee
The variable $\psi$ must have a periodicity of $2\pi$ in order for the
solution to be geometrically smooth at the Taub-NUT centers $\vec x =
\vec x_j$ where the fiber along the $\psi$ direction degenerates.
Similar formulas hold for the $TN_{\hat{k}}$ factors as well, with
$\tilde{L} \equiv \alpha'/L$ playing the role of $R_{11}$ and the
parameters $\vec{x}_{j}$ being replaced by $\vec{\hat{x}}_{i}$.

The positions of the Taub-NUT centers $\vec{\hat{x}}_i$ and $\vec x_j$ will be intepreted as parameterizing the FI and the mass parameters, respectively, and are related to the positions of the NS5 and D5-branes as was given in  (\ref{mass}) and (\ref{FI}). By relating the spectrum of BPS objects, they can be mapped to the following relation in M-theory.
\beq
\vec{m}_{j} &=& \frac{R_{11}\vec{x}_{j}}{2\pi l_{p}^{3}} \\ 
\vec{\xi}_{i} &=& \frac{\tilde{L} \vec{\hat{x}}_{i}}{2\pi l_{p}^{3}}  \ . 
\eeq

\begin{figure}
\begin{center}
\includegraphics[scale=1]{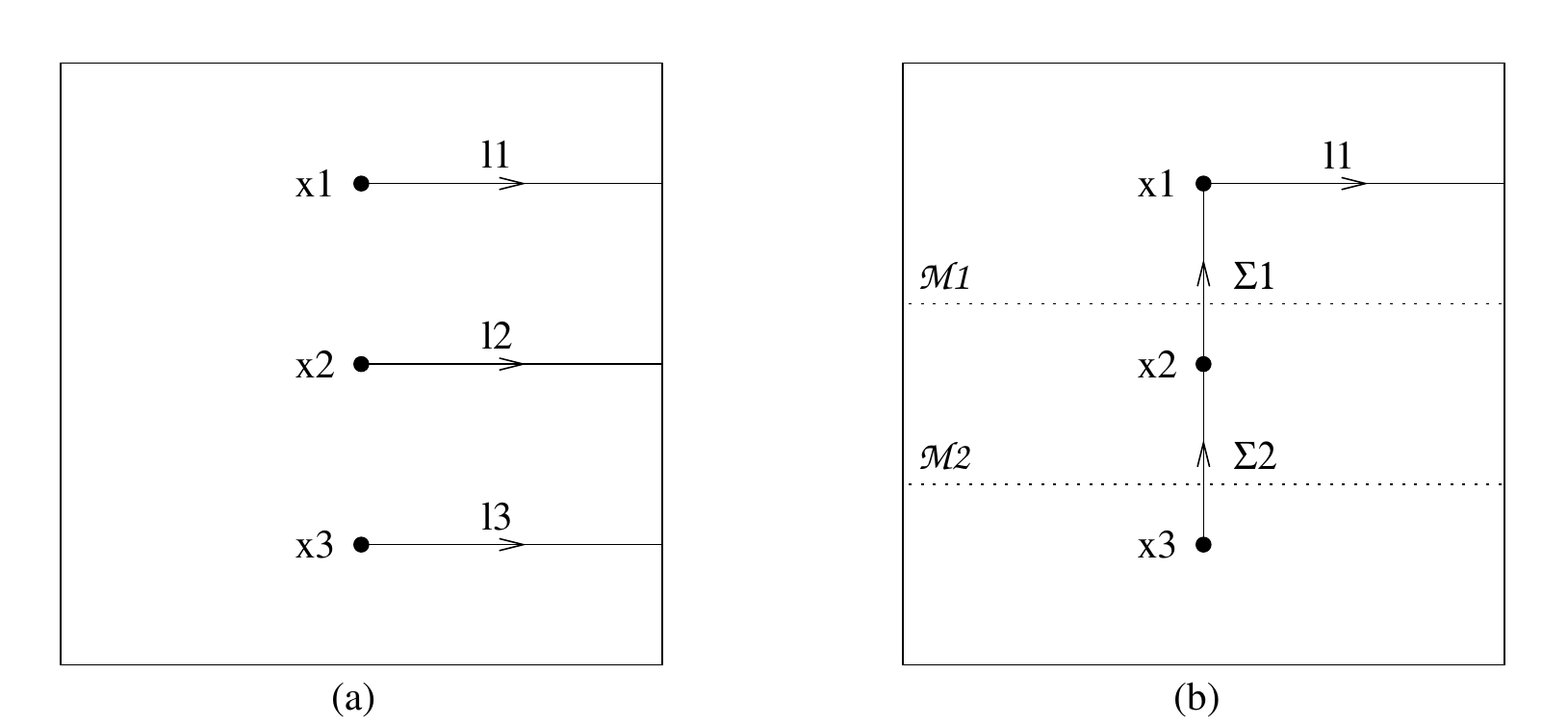}
\caption{In figure (a) the $l_{i}$ are semi-infinite lines in $\mathbb{R}^{3}$ with an $S^{1}$ fiber over each point.  These are the generators of the non-compact homoogy.  The compact homology, figure (b), is formed by fibering the $\psi$ circle over lines between the Taub-NUT centers.  These are homologically equivalent to the differences $l_{i+1}-l_{i}$.  A similar figure appears in \cite{Witten:2009xu}.}
\label{tnlines}
\end{center}
\end{figure}

The $k$-centered Taub-NUT geometry has $k$ normalizable anti-self-dual
2 forms of which $k-1$ is Poincare dual to compact 2-cycles. They can
be parameterized as
\be w_j = d \lambda_j \ee
with the one form $\lambda_j$ given by 
\be \lambda_{j} = \frac{1}{4\pi} \left( 2f_{j}\left(d\psi -\frac{1}{2}\mathbf{w} \right)-\sigma_{j}\right)
\ee
where 
\be \sigma_{j} =\cos\theta_{j}d\phi_{j}\ee
is the potential for the volume form of a unit sphere\footnote{Of course this expression is only valid in certain coordinate patches.}
\be d \Omega_j = d \sigma_j \ee
centered at $\vec x_j$ in $R^3$, and we have defined
\beq
f_{j}&=&-\frac{R_{11}}{2|\vec{x}-\vec{x}_{j}| \left(1+\sum_{\sigma}\frac{R_{11}}{2|\vec{x}-\vec{x}_{\sigma}|}\right)}  \ , 
\eeq
and  the 1-form $\mathbf{w}$ is written in  terms of $\sigma_j$'s as 
\be \mathbf{w}\equiv\sum_{j}\sigma_{j} \ . \ee
These self-dual two forms are normalized so that 
\be \label{wedge}
\int_{TN_{k}} w_{m}\wedge w_{j} = \delta_{mj} \ . 
\ee
It is straight forward then to parameterize the self-dual 4-forms on
$TN_{\hat k} \times TN_k$ as
\be G_4 = (2 \pi l_p)^3 M_i^j \hat w^i \wedge w_j \ee
where $M_i^j$ is a dimensionless numerical coefficient. There are
$\hat k \times k$ independent components in $M_i^j$ corresponding to
$\hat k$ and $k$ linearly independent 2-forms on $TN_{\hat k}$ and
$TN_k$, but they are subject to flux qunatization and boundary
conditions. In order to understand these constraints, it is useful to
visualize the 2-cycles on multi-centered Taub-NUT geometry which we
illustrate in figure \ref{tnlines}.

We begin by noting that there exists a collection of 2-cycles which we
denote $l_i$ which is a semi-infinite line on $R^3$ base of Taub-NUT
geometry with $S^1$ fiber over each point. This cycle is smooth at
each of the Taub-NUT centers labeled as $x_i$.  The period of the two froms $w_j$ on the $l_i$ cycles takes a simple form
\be \int_{l_i} w_j = \delta_{ij} \ . \ee
An alternative scheme to identify the homology cycles is to define $k-1$ cycles
\be \Sigma_i = l_{i+1} - l_i, \qquad i = 1 \ldots k-1 \ee
as is illustrated in figure \ref{tnlines}.b. These cycles are
compact. Similarly, there are $\hat k-1$ cycles $\hat \Sigma_i$ on
$TN_{\hat k}$.  There are therefore $(\hat k -1) \times (k-1)$ compact
4 cycles in $TN_{\hat k} \times TN_k$.

Clearly, the period of $G_4$ on $\hat \Sigma_i \times \Sigma_j$ must
be quantized. That is
\be P_{i}^{j}= {1 \over (2 \pi l_p)^3} \int_{\hat \Sigma_i \times \Sigma_j} G_4^{SD} = \mathbb{Z} \ . \ee
This constrains $(\hat k-1) \times (k-1)$ components of $M_{ij}$. The
remaining $\hat k + k -1$ components can only be characterized in a
gauge invariant manner as a period over a non-compact cycle. Without
loss of generality, we can take that cycle to be $l_1$. Let us define
\beq 
\hat q_i &=& {1 \over (2 \pi l_p)^3} \int_{\hat \Sigma_i \times l_1} G_4^{SD}, \qquad i = 1 \ldots \hat k - 1\\
 q^j &=&  {1 \over (2 \pi l_p)^3}  \int_{\hat l_1 \times \Sigma_j} G_4^{SD}, \qquad j = 1 \ldots k-1\\
r & = &   {1 \over (2 \pi l_p)^3}  \int_{\hat l_1 \times l_1} G_4^{SD} \ . 
\eeq
The set of parameters $(P_{i}^{j}, \hat q_i, q^j, r)$ completely specify
all components of $M_{ij}$.
\be M_{i}^{j} = r + \sum_{i'=1}^{i-1} \hat q_{i'}  + \sum_{j'=1}^{j-1}  q^{j'} + 
\sum_{i'=1}^{i-1}  \sum_{j'=1}^{j-1}  P_{i'}^{j'} \ . \ee
To the extent that $\hat l_1$ and $l_1$ are non-compact, there is no
sense in which $\hat q_i$, $q^j$, or $r$ are to be quantized. There is,
however, a meaning which can be attibuted to the fractional part of
$\hat q_i$, $q^j$, and $r$. The pullback of the 3-form potential
\beq  \hat \gamma_i &=&  {1 \over (2 \pi l_p)^3}  \int_{\hat \Sigma_i \times \partial l_1} C_3^{SD} \simeq \hat q_i \mod \mathbb{Z}  \label{gammabc} \\
\gamma^j &=&  {1 \over (2 \pi l_p)^3}  \int_{\partial \hat l_1 \times \Sigma_j} C_3^{SD} \simeq  q_j \mod \mathbb{Z}  \\
\gamma & = &  {1 \over (2 \pi l_p)^3}  \int_{\partial \hat l_1 \times l_1} C_3^{SD} \simeq 
 r \mod \mathbb{Z} 
\eeq
has the same fractional parts as $\hat q_i$, $q^j$, and $r$ up to
large gauge transformation which can shift their values by integer
amounts.\footnote{It should be noted that roughly speaking $M$
  corresponds to $l$ and $\gamma$ corresponds to $l-\alpha$ in the
  parameterization of (\ref{ahhoansatz}).}

It is convenient to define
\beq \hat b_i &=& \sum_{i'=1}^{i-1} \hat \gamma_i, \qquad i=1 \ldots \hat k \\
b^j &=& \sum_{j'=1}^{j-1}  \gamma_j,\qquad i=1 \ldots k
\eeq
which can be arranged, after large gauge transformation and
permutations, to satisfy
\be 
0 = \hat{b}_{\hat{k}} \le \hat{b}_{\hat{k}-1} \le .... \le \hat{b}_{2} \le \hat{b}_{1} \le 1 \qquad   0\le \gamma \le 1\ .  \label{order}
\ee
As is evident from the notation, $\hat{b}_{i}$$\,\left(b^{j}\right)$
will be identified with the $NS5$$\,\left(D5\right)$ brane position
in, for instance, equation (\ref{gvb}). Also, note that we will not
need to impose any condition on $b_{j}$.  In this setup, we can write
\be M_{i}^{j}  = L_{i}^{j} + \hat b_i + b^{j} + \gamma \ee
where $L_{i}^{j}$ consists of $\hat k \times k$ integer data and $\hat
k + k - 1$ continuous data $\hat{b}_i$, $b^j$, and $\gamma$ consisting
of real numbers between 0 and 1, respecting the constraint
(\ref{order}).  Roughly sepaking, these data can be thought of as
arising from separating the $\hat q_i$, $q^j$, and $r$ into their
integer part and the fractional parts.  Notice, however, that at this
point there is an ambiguity in specifying $L_{i}^{j}$ since the large
gauge transform (\ref{lgauge}) mixes $L_{i}^{j}$ and $b^{j}$ while
keeping $M_{i}^{j}$ invariant.

Fortunately, the ambiguity in $L_{i}^{j}$ is rendered harmless in the
limit that $TN_{\hat k} \times TN_k$ degenerates to $ALE_{\hat k}
\times TN_k$. The key difference between $TN_{\hat k}$ and $ALE_{\hat
  k}$ is the fact that one linear combination of the $\hat k$
self-dual 2-forms
\be \hat \Xi  = \sum_{i=1}^{\hat k} \hat{w}^i \rightarrow 0 \ee
no longer exists as an element of cohomology.  This implies that
$G_4^{SD}$ is independent of $\gamma$ and $b^j$. Hence, the physics is
unaffected by $L_{i}^{j}\rightarrow L_{i}^{j} +p^{j}$, even for fixed
arbitrary $b^{j}$.  In other words, the counting of parameters is
reduced to $L_{i}^{j}$ having $\hat k \times k - k$ distinct discrete
parameters and $M_{i}^{j}$ having $\hat k - 1$ additional continuous
parameters $\hat b_i$. The ambiguity/redudnancy in $L_{i}^{j}$ can be
thought of as the manifestation of (\ref{lgauge}) we encountered in
the previous section. The relations (\ref{pimap1})--(\ref{pimap2})
were derived for $L_{i}^{j}$ with ambiguity (\ref{lgauge}) fixed to
satisfy (\ref{branegauge}). Since $M_{i}^{j}$ decouples from $\gamma$
and $b^{j}$, we can freely adjust $L_{i}^{j}$ to have the same
structure as what we found in the previous section. This provides
compelling reason to identfy the $L_{i}^{j}$ found here as a data of
the gravity solution to the $L_{i}^{j}$ introduced in the previous
section as a data characterizing the brane configuration. In the
following section, we will provide further support for this
identification by relating these discrete parameters to Page charges
in type IIA supergravity.

\section{IIA Description}
\label{IIA}

In the previous section, we consiered a large class of BPS
supergravity backgrouds for M-theory on $TN_{\hat k} \times TN_{k}$
with qunatized fluxes as well as their $ALE_{\hat k} \times TN_k$
limit. In this section, we will consider the type IIA reduction of
these backgrounds and relate the parameters to type IIB brane
construction parameters considered earlier.

The IIA reduction of the ansatz (\ref{ansatz}) can be written in the form
\begin{eqnarray}
\label{GeneralIIARed}
A_{3}&=&-H^{-1}dt\wedge dx_{1}\wedge dx_{2}+(2 \pi l_p)^3 M_{ij}\hat{w}_{i}\wedge \lambda_j \\ \nonumber
B_{2}&=&-\frac{1}{2\pi R_{11}}(2 \pi l_p)^3 M_{ij}\hat{w}_{i}f_{j}-\frac{2}{R_{11}}(2 \pi l_p)^3  \hat{\beta}_{i}\hat{w}_{i} \\ \nonumber
e^{\phi}&=&H^{1/4}V^{-3/4} \\ \nonumber
A_{1}&=&-\frac{R_{11}}{2} \sum_{j} \sigma_{j} 
\end{eqnarray}
where we have written the potential for the IIA form fields explicitly
to match the boundary condition (\ref{gammabc}).  We are also working
in the $ALE_{\hat k} \times TN_k$ limit where some of the components
of $M_{ij}$ have decoupled.  In the IIA descriptions, the branes are
oriented as follows.
\be
\begin{tabular}{c||cccccccccc}
       & 0& 1  & 2& 3& 4& 5& 6& 7& 8& 9 \\
       \hline
NS5 & $\circ$ &  $\circ$ & $\circ$ &  $\circ$ &  $\circ$  &$\circ$   &   &   &   &     \\
D6 & $\circ$ & $\circ$  &$\circ$& & &    & $\circ$  &$\circ$ &$\circ$ &$\circ$     \\
D2 & $\circ$& $\circ$& $\circ$& &  &  &  &  &   &
\end{tabular} \label{orientation}
\ee

One way in which we can explicitly relate the $L_{i}^{j}$ which
appeared in the context of gravity solution in the previous section to
the $L_{i}^{j}$ from the brane construction is to examine the magnetic
flux on $k$ D6-brane extended along the $ALE_{\hat k}$ parameterized
by the (6789) coordinates, and localized along the 345 coordinates in
the dimensional reduction of M-theory to IIA. The presence of a D3
segment stretched between the $i$-th NS5-brane and the $j$-th D5 brane
in the IIB picture translates to the presence of magnetic flux on the
D6 world volume which can be read off from the supergravity solution.

First, let us recall that the gauge invariant $U(1)$ world volume
gauge field on the $j$-th D6 brane can be read of using formulas in
\cite{Imamura:1997ss,Dasgupta:2003us} as follows:
\be {\cal F}^j = B + 2 \pi\alpha' F^j = - {1 \over 2 \pi R_{11}} \int_{l_j} G_4^{SD} =  -(2\pi l_{s})^{2} M_i^j \hat \omega_i
\ee
Here, $B$ is the induced NSNS 2-form on the D6-brane world volume in
the probe approximation. That is, it should be the $B$-field evaluated
at $\vec x$ at infinity.  From (\ref{GeneralIIARed}) we can read off
\be B = -(2\pi l_{s})^{2} \sum_{i} \hat b_i \hat w_i \ . \ee
Combining these expressions, we find that
\be F^j = -2\pi \sum_{i} (M_i^j-\hat b_i) \hat \omega_i = -2\pi \sum_{i} L_i^j \hat \omega_i \ . \ee
When this field strength is dimensionally reduced on $\hat \psi$, we
obtain a $U(1)$ gauge field on an $R^3$ with $\hat k$ marked points
indexed by $i$ on the $R^3$ base space of $ALE_{\hat k}$. The world
volume gauge field becomes that of $L_i^j$ units of magnetic charge
localized at the $i$-th point on the $j$-th D6-brane. This is
precisely what one expects form the T-dual of a D3-brane streteched
between the $i$-th NS5-brane and the $j$-th D5-brane in the type IIB
picture. In the $ALE_{\hat k} \times TN_k$ limit of $TN_{\hat k}
\times TN_k$, the precise value for the ``center of mass'' $\sum_i
L_i^j$ to assign is ambiguous, corresponding to the physical ambiguity
discussed in the context of brane configuration in
(\ref{lgauge}). This ambiguity manifests itself as ambiguity with
respect to large gauge transformation on the supergravity side.

Another quantity that is interesting to compute in the IIA description
to clarify the mapping of parameters under gauge gravity
correspondence is the Page charge. Page charge is one of three
distinct notion of charges identified by Marolf \cite{Marolf:2000cb}
in background with fluxes and has the property of being localized,
conserved, but not invariant under large gauge transformations in some
cases. They are defined as periods of Page flux over cycles containing
the charge source. A convenient reference for the Page fluxes and
various subtle supergravity conventions we follow can be found in
appendix B of \cite{Aharony:2009fc}.

The D6 Page charge is simply 
\be Q_6^{Page} = \frac{1}{2\pi R_{11}} \int F_2 = k \ee
and counts the number of D6-branes. 

Similarly, we define the D4-Page charge associated to D4-branes
warpped on each of the $(\hat k-1)$ 2-cycles $\hat \Sigma_i$ in
$ALE_{\hat k}$ by integrating the D4 Page flux over the cycle $\hat
{\cal M}_i$ orthogonal to $\hat \Sigma_i$ as was defined in figure
\ref{tnlines}.b.
\be Q_{4i}^{Page} =   {1 \over (2 \pi l_p)^3} \int_{\hat {\cal M}_i \times S^2} (-\tilde F_4 - B_2 \wedge F_2)  =  -{1 \over (2 \pi l_p)^3} \int_{\hat {\cal M}_i \times S^2} d(A_3 + B_2 \wedge A_1)
\ee
where $S_2$ is the sphere on $R^3$ containing all the centers of the
$TN_k$ viewed as an $S^1$ fibration over $R^3$.  Substituing the type
IIA background, we find that these D4 Page charges evaluates to
\be
Q_{4i}^{Page}=\frac{1}{2}\left(\sum_{n=i+1}^{\hat{k}} \hat{l}_{n}-\sum_{n=1}^{i} \hat{l}_{n}\right) +\frac{1}{2\hat{k}} \left(2 i-\hat{k}\right) \sum_{n=1}^{\hat{k}-1} n\, k_{n} ,\qquad i = 1 \ldots (\hat k-1) \ . 
\ee
In performing this calculation, the following formula for computing the intersection form on $ALE_{\hat k}$ is useful.
\be \int_{ALE_{\hat k}} \hat w_i \wedge \hat w_j = \delta_{ij} - {1 \over \hat{k}} \ . \ee
Also, it should be noted that $Q_{4i}^{Page}$ is only defined modulo
$k$ due to gauge ambiguities.

Similar computation of the D2 Page charge gives
\beq
\label{q2Page}
Q_{2}^{Page} &=& Q_{M2}+\frac{1}{2} \sum_{i,j} \left(L_{i}^{j}-\frac{1}{\hat{k}}\sum_{i}L_{i}^{j}\right)^{2} \ . 
\eeq

The fact that $Q_2^{Page}$ and $Q_{4i}^{Page}$ are all expressed in
terms of discrete quantities is consistent with the expectation that
the Page charges are discrete, conserved quantities. It may come as a
bit of a surprise that these quantities do not take integer values
themselves. Page charges, however, have been known to contain
anomalous additive contributions, such as the Freed-Witten anomaly,
that can shift their values away from strictly integral values.  It
was in fact anticipated that the $Q_2^{Page}$ and $Q_4^{Page}$ takes
on a specific form (\ref{pageanomaly}) and (\ref{pageanomaly2}) for
the model whose quiver diagram is illustrated in figure
\ref{disflavors} with $\hat k=2$. Using the data for $L_i^j$ which we
worked out in (\ref{lLL}), we find that (\ref{pageanomaly}) and
(\ref{pageanomaly2}) are reproduced precisely.  Although we have
computed the total Page charge for the resolved theory, we expect the
same result for the orbifold limit since Page charges, being discrete,
are invariant under continuous deformations.

Another useful quantity that we can compute to characterize the type
IIA background is the Maxwell charge, which is gauge invariant and
conserved but is not invariant under deformation or quantized. It is
given by
\be Q_2^{Maxwell} = {1 \over (2 \pi l_s)^5 g_s} \int_{ALE_{\hat k} \times S^2} * \tilde F \ee
and evaluates to
\be Q_2^{Maxwell} = Q_{M2} + {1 \over 2} \sum_{i,j}  \left( \hat L_i^j - \hat \beta_i - {1 \over \hat k} \sum_m  (L_m^j -  \hat \beta_m) \right)^2 \ee
which also be written in the form
\beq 
Q_2^{Maxwell} &=& Q_2^{Page} + Q_{4m}^{Page} \int_{ \hat  \Sigma_m} B +  {k \over 2} \int_{ALE_{\hat k}}  B \wedge B \cr
&=&
Q_2^{Page} +    Q_{4i}^{Page}  \hat \gamma_{i} + {k \over 2} ({\cal I}^{-1})_{i i'} \hat \gamma_i  \hat \gamma_{i'}
\eeq
where ${\cal I}_{ii'}$ is the $(\hat k-1) \times (\hat k -1)$ intersection matrix (\ref{ALEintersect}) on $ALE_{\hat k}$:  
\be {\cal I}_{ij} = \left(
\begin{array}{ccccc}
2 & -1 & \\
-1 & 2 & -1 \\
& \ddots &\ddots & \ddots \\
 & & -1 & -2 & -1 \\
 & &    & -1 & 2
\end{array}\right) \label{ALEintersect}\ee
It is easy to interpret the Maxwell charge as the net D2 charge
induced by the $B$-field being pulled back on the $Q_{4i}^{Page}$ D4
and $k$ D6 branes when written in this form. In this sense, the Page
charges are counting the effective number of branes that are present
albeit with some shift.

In the $\hat k=2$ limit, this reduces to 
\be Q_2^{Maxwell} = Q_2^{Page} + Q_{4}^{Page} \gamma_1 + {k \over 4} \gamma_1^2 \ee
and agrees with the results previously reported in
\cite{Aharony:2009fc,Cottrell:2015jra} when $\gamma_1$ is identified
as $b_\infty$.

Another important quantity which is useful for understanding the
supergravity dual is the D2 brane charge and the bulk charge. The bulk
charge refers to the contribution to the Maxwell charge arising from
the $G_4 \wedge G_4$ term in the M-theory equation
\be  d * G_4 - \frac{1}{2}G_4 \wedge G_4 = (2\pi l_{p})^{6}* j_{M2}^{brane} \ee
in M-theory and the IIA reduction thereof. 

So
\be Q_2^{Bulk} = \frac{1}{2(2\pi l_p)^{6}} \int_{ALE_{\hat k} \times TN_k} G_4 \wedge G_4 = {1 \over 2} \sum_{i,j}  \left( \hat L_i^j - \hat \beta_i - {1 \over \hat k} \sum_m  (L_m^j -  \hat \beta_m) \right)^2  \ . \ee
The brane charge is the contribution from the source term and simply
evaluates to
\be  Q_2^{brane} = \int_{ALE_{\hat k} \times TN_k} *j_{M2}^{brane} = Q_{M2} \ . \label{Q2brane} \ee
Here, we have formulate these quantities in the language of M-theory
but they reduce naturally to type IIA and evaluates to the same
value. This analysis was for the case where the $ALE_{\hat k}$ and
$TN_{k}$ were completely resolved and as such, the only possible
localized source contributing to $*j_{M2}$ is the isolated M2/D2
brane. $Q_2^{brane}= Q_{M2}$ is therefore naturally an integer valued
quantity.

That $Q_2^{brane}$ is strictly integral and positive definite is
curiously at odds with the earlier claim (\ref{sugbnd}).  It turns
out that this apparent mismatch has some very interesting physical
origin. 

First, let us note that the self-dual 4-form can be decomposed into
compact and non-compact part
\be G_4 = G_4^{compact} + G_4^{noncompact} \ee
where 
\be G_4^{non-compact} =  {(2\pi l_{p})^{3} \over k} (\sum_{j=1}^k M_i^j) \hat w^i \wedge (\sum_j w_j) \ee
and so
\beq 
G_4^{compact} &=& G_4 - G_4^{non-compact}  \\ 
&\equiv& (2\pi l_{p})^{3} \sum_{ij} \hat{M}_{i}^{j}
\eeq
where $G_4^{compact}$ is the part which has non-trivial period on
$\Sigma_j$'s whereas
\be \int_{\Sigma_j} G_4^{non-compact} = 0 \ . \ee
The compact and non-compact components are orthognoal, so that
\be \frac{1}{2(2\pi l_{p})^{3}} \int_{ALE_{\hat k} \times TN_k} G_4 \wedge G_4 
= Q_2^{compact} + Q_2^{non-compact} \ee
with
\beq
Q_2^{compact} &=& \frac{1}{2(2\pi l_{p})^{3}}\int_{ALE_{\hat k} \times TN_k} G_4^{compact} \wedge G_4^{compact} \\
Q_2^{non-compact} & = & 
\frac{1}{2(2\pi l_{p})^{3}}\int_{ALE_{\hat k} \times TN_k} G_4^{non-compact} \wedge G_4^{non-compact} \ . 
\eeq
The point of decomposing $G_4^{SD}$ into compact and non-compact
components is that they behave very differently in the orbifold limit
of $TN_k$. The non-compact component has a smooth limit. Nothing
special happens to the non-compact component in taking the orbifold
limit.

Not too surprisingly, the compact component of the bulk charge
exhibits more intricate behavior in the orbifold limit.  The compact
component of $G_4 \wedge G_4$, in fact, degenerates to a delta
function in the orbifold limit. One way to see this explicitly in a
simple example is to look at the compact self-dual 2-form on
Eguchi-Hanson space in the orbifold limit given, for instance, in
equation (B.4) of \cite{Bertolini:2001ma}.

Upon explicit evaluation, one finds
\beq
Q_{2}^{compact}&=& \frac{1}{2} \left(\sum_{i,j} (\hat{M}_{i}^{j})^{2} -\frac{1}{\hat{k}}\sum_{i}\sum_{n,m}\hat{M}_{n}^{j} \hat{M}_{m}^{j}\right)\\
Q_{2}^{non-compact}&=& \frac{1}{2k} \left(\sum_{i,j,k} M_{i}^{j}M_{i}^{k}-\frac{1}{\hat{k}} \left(\sum_{i,j}M_{i}^{j}\right)^{2}\right) \ . 
\eeq
To the extent that $Q_2^{compact}$ is delta-function supported at the
tip of the orbifold, it behaves in many way like a brane
charge. Suppose for the sake of argument we consider the combination
\be Q_2^{brane} + Q_2^{compact} = Q_{M2}+\frac{1}{2} \sum_{i,j} \left(L_{i}^{j} +l^{j}/\hat{k}-\frac{1}{k}\sum_{j}\left(L_{i}^{j}+l^{j}/\hat{k}\right)\right)^{2} \ . 
\ee
Now, using (\ref{q2Page}), one finds 
\beq Q_2^{brane} + Q_2^{compact} &=&Q_{2}^{Page}-\frac{1}{2k} \sum_{i}\left(\sum_{j}\left(L_{i}^{j}+\frac{l^{j}}{\hat{k}}\right)\right)^{2}
\eeq
which looks somewhat complicated. However, when restricted to $\hat
k=2$ and $k_{2}=k$, which was the case considered in
\cite{Aharony:2009fc}, this equation simplifies to
\be Q_2^{brane} + Q_2^{compact} = Q_2^{Page} - {(Q_4^{Page})^2 \over k} \label{branecompact} \ee
which now can be seen as having the same form in the right hand side
as (\ref{sugbnd}).

We have therefore gained a useful perspective on why (\ref{Q2brane})
and (\ref{sugbnd}) gave seemingly different results. The two main
difference is that 1) the earlier expression (\ref{sugbnd}) included
contribution from $G_4^{compact}$, and 2) was expressed in terms of
$Q_2^{Page}$ which differ from $Q_{M2}$ as is given in (\ref{q2Page}).

In fact, when the $ALE_{\hat k} \times TN_k$ is resolved,
(\ref{q2Page}) implies that
\be Q_{M2} = Q_2^{Page} - \frac{1}{2} \sum_{i,j} \left(L_{i}^{j}-\frac{1}{\hat{k}}\sum_{i}L_{i}^{j}\right)^{2} > 0 \label{strong} \ee
gives rise to a {\it stronger} condition for perserving supersymmetry
than the condition that (\ref{sugbnd}) be positive. 

The only remaining issue is whether the bound (\ref{strong}) can be
violated if large localized charge is present due to fluxes threading
the collapsed cycles in the orbifold limit to ensure in
(\ref{branecompact}) is positive. This is a subtle issue of topology
change and is beyond the scope of classical gravity analysis. It would
be very interesting to understand the local behavior of this system
around this transition point, as this issue lies at the hart of how
the meeting of Higgs and the Coulomb branch is captured in
gauge/gravity duality.

\section{Conclusions}

In this article, we described an explicit construction of gravity dual
of ${\cal N}=4$ $\prod_{i=1}^{\hat k} U(N_i)$ quiver gauge theory with
$k_i$ fundamentals charged under $U(N_i)$ and bi-fundamentals,
generalizing the earlier construction of \cite{Aharony:2009fc}. The
supergravity description we find captures the full renormalization
group flow starting from the UV point in 2+1 dimensions, at least in
the regime where the gravity approximation is reliable. Our
construction consisted primarily of generalizing the structure of
$ALE_2 \times TN_k$ used as the starting point in
\cite{Aharony:2009fc} to $ALE_{\hat k} \times TN_k$.  We further
resolved the $ALE_{\hat k}$ and $TN_{k}$ geometry to be completely
regular, and allowed fluxes to thread through the resulting compact
four cycles. These fluxes turns out to encode the discrete data
characterizing $N_i$, $k_i$, and the choice of vacua among the set of
discrete, degenerate, choices.  The mapping between the fluxes and
these discrete data is somewhat cumbersome, but is accessible. The
results are summarized in (\ref{pimap1})--(\ref{pimap2}).

The quantization of fluxes and charges takes a relatively simple form
when formulated as an exercise in M-theory on $R^{1,2} \times
ALE_{\hat k} \times TN_k$ when the orbifold singularity is completely
resolved. All the localized M2 sources corresonds to physical
M2-branes, and as such have positive definite, integer quantized
values for supersymmetric solutions. This is somewhat at odds with the
previous finding of type IIA brane charge (\ref{Q2b}) whose values
were only quantized in units of $1/k$. We found a gratifying
resolution to this apparent discrepancy. Upon taking the orbifold
limit, the fluxes threading the compact cycles approach an effective
delta-function source at the orbifold fixed point. These sources can
carry charges that are quantized in units of $1/k$ relative to the M2
charge. In essence, the flux thorugh the compact cycle transmutes to
discrete torsion like in \cite{Sethi:1998zk,Bergman:2009zh}.

What is interesting about the orbifold limit of our construction based
on resolved $ALE_{\hat k} \times TN_k$ is that it should correspond to
the point on the moduli-space where the Coulomb branch and the Higgs
branch meet. It would be very interesting to understand this
transition concretely from the bulk description and account in detail
features such as the dimension of Coulomb branch.

We seem to be finding, however, that much of the interesting field
theory dynamics requires understanding the stringy resolution of
classical gravity. For models whose parameters are outside the range
(\ref{gwgood}), one must resolve the enhancons in order to extract
meaningful physics. Even for models whose parameters are contained in
the (\ref{gwgood}), string corrections become important when
approaching the threashold of supersymmetry. In fact, the discrepancy
between (\ref{ftsusy}) and (\ref{sugbnd}) can be attributed entirely
to the stringy uncertainty as was outlined in the
introduction. Similar stringy corrections also arise in ABJM model as
was shown in \cite{Drukker:2010nc}.  If the nature of stringy
corrections can be classified on the gravity side, perhaps one can use
gauge gravity correspondence to also classify the stringy effects
using knwon exact results such as the ones obtained via localization
\cite{Nishioka:2011dq,Assel:2012cp}.

\begin{figure}
\includegraphics[width=\hsize]{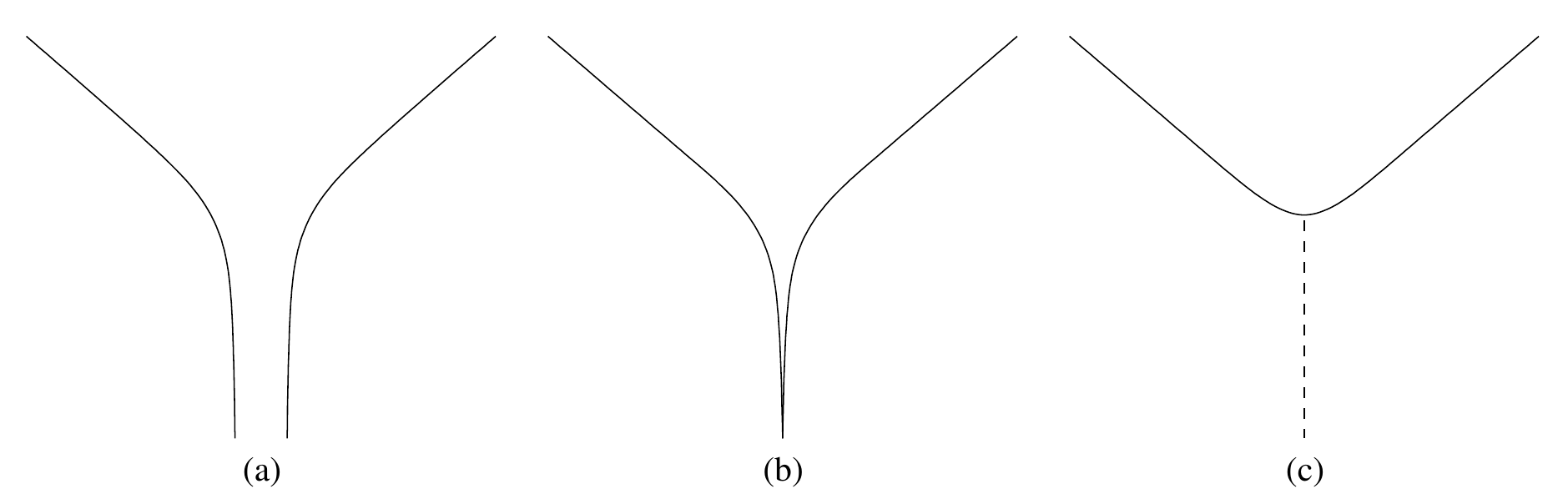}
\caption{Schematic illustration of the class of geometries which consists of asymptotic region and (a) a macroscopic $AdS_4$ throat, (b)  a small $AdS_4$ throat, and (c) semiclassicaly smooth capped of geometry, potentially accompanied by strucures only visible when string corrections are included. \label{figgeom}}
\end{figure}

Perhaps one of the most profound implication of the string correction
is the status of gravity solution for parameters corresponding the
edge of the parabola illustrated in figure \ref{akiplot}.  For
solutions in the interior of the parabola in the enhancon free region
(\ref{gwgood}), one expects an $AdS_4$ throat but at the edge of the
parabola, that throat disappears and one seemingly obtains a regular
geometry capped off near the origin.  A large class of ``regular''
solutions of this type were constructed in a impressive body of work
by Cvetic, Gibbons, Lu, and Pope starting with \cite{Cvetic:2000db}
and reviewed in \cite{Cvetic:2002kn}. The singularity free regular
solution being constructed in these works primarily involved sitting
at the edge of the parabola. Attempts to explore dynamical
supersymmetry breaking along the lines of
\cite{Cottrell:2013asa,Giecold:2013pza} consited of exploring the
region near the edge of the parabola. The picture emerging from the
consideration of string corrections, however, is the fact that the
concept of tuning of parameter to sit at the edge of the parabola
itself is subject to correction. This is because as the edge of the
parabola is approached from the inside, the $AdS_4$ throat becomes
highly curved, and the semi-classical formula relating the radius to
charges starts to receive corrections. The issue here stems largely
from if/whether one achieves the smooth, capped off geometry in the
limit of vanishing $AdS_4$ radius. Recently, in a very interesting
series of papers, it was argued that a tip of the $SL(2)/U(1)$ cigar
receives $\alpha'$ corrections which are non-perturbative dramatically
alterning the semi-classical intuition that the tip of the cigar is
smooth \cite{Giveon:2012kp,Giveon:2013ica,Giveon:2015cma}. One way to
understand the physics of stringy corrections in their context was to
say that translation invariance in the winding space was broken by
condensation of winding modes. Something very similar could be
happening near the edge of the parabola as is illustrated in figure
\ref{figgeom}.  There are indeed compelling evidence that translation
invariance in the periodic direction T-dual to $\hat \psi$ is broken
in the explicit construction of supergravity duals of the
superconformal IR fixed point for theories in the ``good'' region
(\ref{gwgood}) \cite{Assel:2011xz,Assel:2012cj}.  Microscopic
corrections to gravity that are not immediately apparent from the
gravity as a low-energy effective field theory is at the heart of the
black hole information paradox, and it is quite interesting to find a
manifestation of such effects even in a highly supersymmetric setup
like the ones considered in this note. These issues will likely
continue to play a role in clarifying what fuzzballs and firewalls
really mean, and will likely also have an impact on how gauge/gravity
correspondences are understood.

\section*{Acknowledgements}
We thank O.~Aharony, S.~Hirano, P.~Ouyang, and J.~Hanson for collaboration on related work, and O.~Aharony, C.~Bachas, E.~Conde-Pena, N.~Itzhaki, M.~Moskovic, A.~Royston for discussions and correspondences.

\bibliography{fi}\bibliographystyle{utphys}

\end{document}